# Two-dimensional Indium Oxide at the Epitaxial Graphene/SiC Interface: Synthesis, Structure, Properties, and Devices

*Furkan Turker[1,2], Bohan Xu[3], Chengye Dong[2,4], Michael Labella III[5], Nadire Nayir[6,7,8], Natalya Sheremetyeva[9], Zachary J. Trdinich[1], Duanchen Zhang[1], Gokay Adabasi[10], Bita Pourbahari[11,12], Wesley E. Auker[5], Ke Wang[5], Mehmet Z. Baykara[10], Vincent Meunier[9], Nabil Bassim[11,12], Adri C.T. van Duin[1,8,9,13], Vincent H. Crespi[1,2,3,4,13], Joshua A. Robinson[1,2,3,4,9,13]\**

[1]Department of Materials Science and Engineering, The Pennsylvania State University
[2]Center for 2-Dimensional and Layered Materials, The Pennsylvania State University
[3]Department of Physics, The Pennsylvania State University
[4]Two-Dimensional Crystal Consortium, The Pennsylvania State University
[5]Materials Research Institute, The Pennsylvania State University
[6]Paul-Drude-Institute for Solid State Electronics, Leibniz Institute within Forschungsverbund Berlin eV
[7]Department of Physics Engineering, Istanbul Technical University
[8]Department of Mechanical Engineering, The Pennsylvania State University
[9]Department of Engineering Science and Mechanics, The Pennsylvania State University
[10]Department of Mechanical Engineering, University of California Merced
[11]Department of Materials Science and Engineering, McMaster University
[12]Canadian Centre for Electron Microscopy, McMaster University
[13]Department of Chemistry, The Pennsylvania State University

*Corresponding author's email: jrobinson@psu.edu

## Abstract

High-quality two-dimensional (2D) dielectrics are crucial for fabricating 2D/3D hybrid vertical electronic devices such as metal-oxide-semiconductor (MOS) based Schottky diodes and hot electron transistors, the production of which is constrained by the scarcity of bulk layered wide bandgap semiconductors. In this research, we present the synthesis of a new 2D dielectric, monolayer $InO_2$, which differs in stoichiometry from its bulk form, over a large area (>300 µm$^2$) by intercalating at the epitaxial graphene (EG)/SiC interface. By adjusting the lateral size of graphene through optical lithography prior to the intercalation, we tune the thickness of $InO_2$ where predominantly (~85%) monolayer $InO_2$ is formed. The preference for monolayer formation of $InO_2$ is explained using ReaxFF reactive molecular dynamics and density functional theory (DFT) calculations. Additionally, the band gap of $InO_2$ is calculated to be 4.1 eV, differing from its bulk form (2.7 eV). Furthermore, MOS-based Schottky diode measurements on $InO_2$ intercalated EG/n-SiC demonstrate that the EG/n-SiC junction transforms from ohmic to a Schottky junction upon intercalation, with a barrier height of 0.87 eV and a rectification ratio of ~$10^5$. These findings introduce a new addition to the 2D dielectric family, showing significant potential for monolayer $InO_2$ to be used as a barrier in vertical electronic devices.

**Keywords:** 2D Indium oxide, intercalation, graphene, heterostructure, vertical Schottky diode

**Introduction**

Confining species at the interface between epitaxial graphene (EG) and silicon carbide (SiC) paved the way for the creation of large-scale, stable monolayer to few-layer metals,[1] metal alloys,[2] and metal compounds.[3,4] The confinement of these metallic structures at this distinctive asymmetric interface has led to the development of several unique properties, including metal-to-semiconductor transitions[5] and superconductivity.[1] This enables opportunities to explore 2D dielectrics, a field constrained by the narrow range of available wide bandgap semiconductors that are bulk and layered. An early, exemplary paper is the encapsulation of 2D-$GaN_x$[4] (noted for its bandgap of ~5 eV) – since then, the intercalation of high bandgap compounds, such as $AlN_x$,[6] $GaN_x$,[7] $InO_x$,[8] $GaO_x$,[3] have been demonstrated.

Dielectric intercalation at the EG/SiC interface offers a foundation for developing 2D/3D hybrid vertical electronic devices such as metal-oxide-semiconductor (MOS)-based Schottky diodes and hot electron transistors (HETs). Previous HET studies utilizing EG/SiC, generally report poor output characteristics and a low on-off ratio due to high leakage current through the as-grown EG/SiC junction.[9] The ohmic behavior of the EG/SiC (0001) interface can be converted into a Schottky junction through hydrogen intercalation.[10] This transformation results from the differences in graphene's work function and the electron affinity of SiC as well as the spontaneous polarization of hexagonal SiC. The reported barrier height at the hydrogen-intercalated EG/SiC (0001) interface varies significantly from 0.8 to 1.6 eV, indicating that the conditions of graphene growth and hydrogen intercalation are crucial parameters for controlling electron transport across the EG/SiC interface[10–12]. An alternative method for the EG/SiC based Schottky diode formation is via dielectric intercalation, which yields a higher threshold voltage ($V_{th}$) and rectification ratio (RR) than H-intercalated.[3,4,8]

Building on these advancements, the formation of 2D indium oxide through intercalation is particularly appealing due to its high bandgap (~2.7 eV) and as indium intercalation over a large area is readily achievable[1]. Common techniques used to produce ultrathin $InO_x$, such as atomic layer deposition (ALD)[13] and printing an oxide skin from liquid metal,[14] typically yield large area coverage but result in poor crystallinity due to low temperature annealing (< 250 °C). Therefore, intercalation at the EG/SiC interface offer a viable approach for achieving highly crystalline, ultrathin $InO_x$ epitaxial to SiC substrate. Based on our knowledge, there is only one study experimentally demonstrating the intercalation 2D $InO_x$ (bilayer) at the EG/SiC interface where $In(CH_3)_3$ and trace amount of impurities ($H_2O$ and $O_2$) in the gas stream were used as In and O precursors via metal organic chemical vapor deposition (MOCVD).[8] Although Schottky barrier formation upon $InO_x$ intercalation was verified via conductive atomic force microscopy (C-AFM), practical device implementation was hindered by the small lateral size of the 2D $InO_x$.

While significant advancements in the lateral uniformity of metals intercalated at the EG/SiC interface to the centimeter scale (excluding step edges of SiC) have been achieved,[1] advancements in compound intercalation have not kept pace. Challenges in large area synthesis persist for oxides and nitrides[4,7] due to quasi-3D growth at the EG/SiC interface, leading to the formation of graphene cracks. A key distinction exists between metal and compound intercalation: metals such as Ga, In, Sn, Ag, and Au exhibit self-limiting growth[1,5] and stabilize at 1-3 layers thick at the EG/SiC interface. On the other hand, compounds, such as $GaN_x$, can exceed 15 layers. This, in turn, causes graphene to crack and restricts the lateral dimensions of 2D $GaN_x$ to just a few microns.[4,15] These observations underscore the importance of developing new synthesis techniques for the compound formation via intercalation.

In this study, we demonstrate that graphene patterning via optical lithography prior to the $InO_x$ formation significantly improves the uniformity of 2D $InO_x$ layers. This method, dubbed selective area confinement heteroepitaxy (SA-CHet), allows for the intercalation of $InO_x$ over areas as large as 300 μm² at the EG/SiC interface. Moreover, using patterned or continuous graphene results in the formation of monolayer $InO_2$ (different stoichiometry than bulk $In_2O_3$) or multilayer $InO_x$, respectively. Using high resolution transmission electron microscopy (TEM) and theoretical calculations via multi-physics simulations that combine density functional theory (DFT) and Reax FF reactive molecular dynamics simulations, we explain the mechanism leading to thickness variation in these experimental setups. We further elucidate the structural and electronic properties of monolayer $InO_2$, such as the phonon and electronic band structure, both theoretically through DFT and experimentally using Raman spectroscopy and electron energy loss spectroscopy (EELS). Finally, given the large lateral size of 2D $InO_2$, a MOS based Schottky diode is fabricated using EG/$InO_2$/n-SiC which exhibits current densities of ~$10^5$ A/cm² and rectification ratios of ~$10^5$.

**Results and Discussion**

$InO_x$ is formed at the EG/SiC interface using continuous (route 1) or patterned EG (route 2) (**Figure 1**). Route 1 employs confinement heteroepitaxy (CHet), wherein continuous EG (1 x 1 cm²) undergoes low-power $O_2$ plasma treatment (50 W) to induce defects, followed by In intercalation, similar to the process described in Ref [1] (Figure 1a, S1).[1] In Route 2, SA-CHet is employed, where as-grown EG is first patterned via optical lithography and etched to create predefined graphene circles with 20 µm diameter before In intercalation (Figure 1b). Since In can intercalate through the graphene edges in the pattern, where defects are more prevalent, the $O_2$ plasma treatment is unnecessary (Figure S2, S3). Post-indium intercalation Raman spectra and mapping, presented in Figure S2, show continuous ultra-low frequency (ULF) metallic In peak (17 cm$^{-1}$), confirming uniform In intercalation.[16] Note that thick In particles form beneath the graphene following intercalation, serving as In sinks during this process (Figure S4b, c). During oxidation, these regions are the first to exhibit graphene cracking, presumably due to high stress, leading to In deintercalation and non-uniform oxide formation (Figure S4d). Importantly, the density of these particles decreases with larger graphene pattern size and diminishes entirely when intercalation is conducted using the whole sample (1x1 cm²). Hence, graphene is initially patterned into large squares (1.5 mm x 1.5 mm) and intercalated with In without plasma treatment (Figure S5). Subsequently, it is further etched into circles with a 20 µm diameter using an additional lithography step before oxidation. Post-etching, optical micrographs and Raman indium peak mapping confirm a smoother surface and uniform intercalation (Figure S4f, g). This method significantly suppresses graphene crack formation during oxidation, as evidenced by scanning electron microscopy (SEM) images showing a smoother surface (Figure S4h). The two-step etching method is not included in Figure 1b for simplicity.

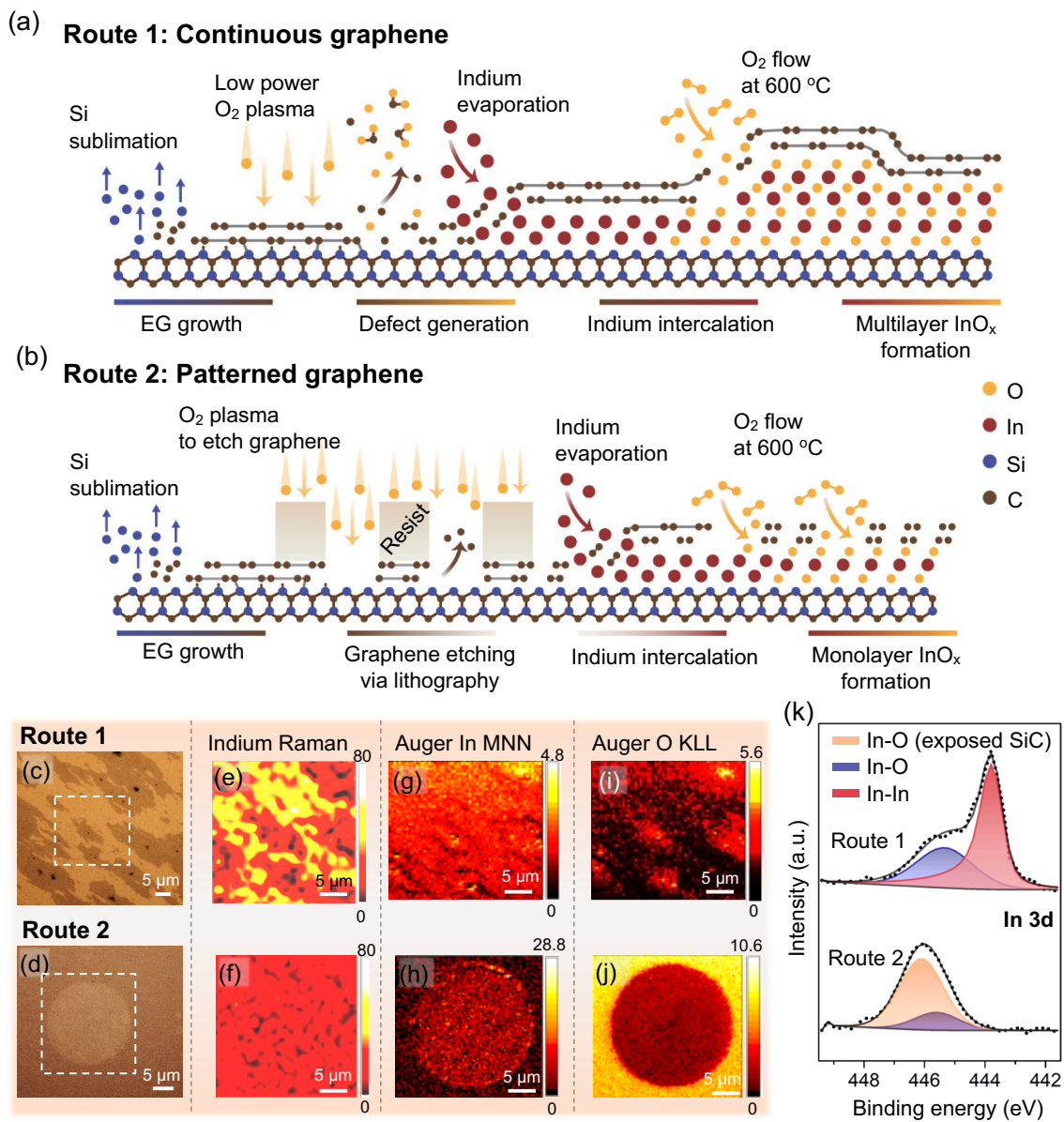

**Figure 1:** Illustration of the routes for the intercalation of 2D $InO_x$ at the EG/SiC interface by using continuous (a) and patterned (b) graphene. Optical micrographs (c, d), Raman maps of metallic indium (e, f), AES In MNN (g, h) and O KLL (I, j) maps, and XPS high resolution In 3d spectra (k) of the indium intercalated samples after oxidation at 600 °C, demonstrating that oxidation with continuous graphene (route 1) yields mixed In/$InO_x$, while fully oxidized indium can be obtained with patterned graphene (route 2).

The oxidation behavior of 2D indium at the EG/SiC interface is dependent on graphene pre-treatment (e.g. continuous or patterned). In both approaches, samples are oxidized at 600°C for 30 minutes under $O_2$ flow. Post-oxidation optical micrograph of the continuous graphene sample (Figure 1c) reveals "bright and dark" contrast regions. Raman mapping (Figure 1e) confirms the bright regions retain metallic In, while Auger electron spectroscopy (AES) mapping (Figure 1i) and X-ray photoemission spectroscopy (XPS) (Figure 1k) indicate incomplete oxidation. Additionally, AES In MNN mapping (Figure 1g) indicates In deintercalation from the EG/SiC interface occurs during oxidation. Conversely, post-oxidation optical micrographs of patterned graphene samples display uniform contrast within the circle (Figure 1d). Raman spectroscopy substantiates the absence of metallic In (Figure 1f), and complete oxidation is confirmed via XPS (Figure 1k). The additional $InO_x$ component in the In 3d spectrum (Figure 1k, orange curve) is due to the X-ray spot size being larger than the circle size is likely to be originated from the $InO_x$ particles on the bare $SiO_x$/SiC surface or $InO_x$ diffused to the SiC substrate during intercalation. Notably, continuous indium and oxygen signals are detected in the AES In MNN and O KLL maps (Figure 1h, j), indicating uniform oxidation. The reduced oxygen count inside the circle compared to the bare SiC region (Figure 1j) is due to oxidation of the SiC surface, as these regions lack graphene coverage. Further, AES point scans inside the circle (Figure S6c) substantiate the successful oxidation of In via graphene patterning. Direct evidence of $InO_x$ intercalation is demonstrated in the cross-sectional scanning TEM (STEM) image, accompanied by energy dispersive X-ray spectroscopy (EDS) elemental maps (**Figure 2a)**.

The lateral size of the graphene layer impacts the thickness of intercalated $InO_x$, with nearly 85% of $InO_x$ formed with patterned EG is monolayer (Figure 2a, c), while continuous graphene layers yield non-uniform $InO_x$ thicknesses of 2-6 layers thick (Figure 2b, c). STEM analysis indicates metallic In intercalates as bilayer (Figure S2e), matching the most stable thickness for In at the EG/SiC interface[1]. Reduction in the In thickness to monolayer during oxidation indicates In deintercalation occurs via the edges of patterned graphene, supported by the observation of bulk $In_2O_3$ particles at the perimeter of the circles (Figure S7). On the other hand, with continuous graphene, In is confined at the interface and can only diffuse laterally beneath EG. This confinement leads to the formation of a broad range of $InO_x$ thicknesses (Figure 2c), including regions without In (Figure 1g).

Graphene defect density impacts the indium oxidation mechanism. We find that defects generated via $O_2$ plasma in continuous graphene are not fully healed during metal intercalation, as the graphene D/G Raman peak ratio is $0.15 \pm 0.04$, whereas the graphene in patterned samples do not exhibit a measurable Raman D peak (Figure S2b). The combination of higher defect density and stress in continuous graphene due to the multilayer $InO_x$ formation can cause the graphene to become more prone to oxidation damage at high temperatures, resulting in "holes" in the graphene. As seen from AES indium mapping (Figure 1g), this leads to indium deintercalation during oxidation which explains the formation of high-density bulk $In_2O_3$ particles on graphene surface and high root mean square roughness (RMS, $R_q = 1.99 \pm 0.25$ nm) (Figure S8). On the other hand, RMS roughness of the patterned graphene surface following the oxidation is 4x lower ($0.52 \pm 0.02$ nm). This, along with the presence of bulk $In_2O_3$ particles observed at the perimeter of the circles verifies that the 2nd layer of indium can diffuse outward during oxidation when the graphene lateral dimensions are on the order of ~20 μm.

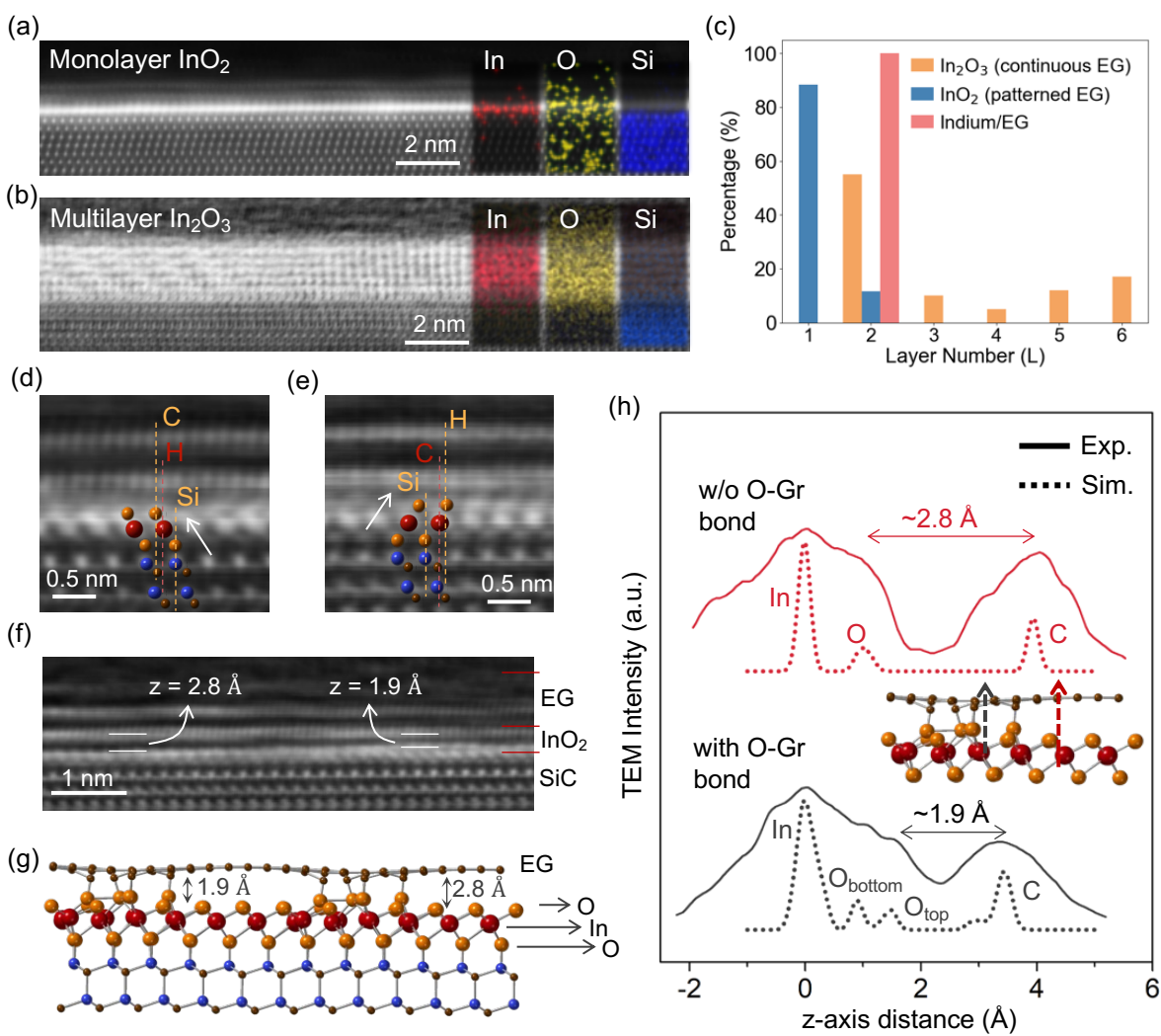

**Figure 2:** Structural analysis of 2D $InO_x$ via cross-sectional HAADF-STEM and DFT. Cross-sectional STEM images and EDS analysis of monolayer $InO_2$ (a) and multilayer $In_2O_3$ (b) formed using patterned and continuous graphene, respectively. Thickness analysis of In and $InO_x$ intercalated graphene (c) measured from STEM images. STEM images of energetically degenerate monolayer $InO_2$ structures (d, e) with Si-H-C and Si-C-H stacking sequences (see supporting information). Si, C, and H in (d) and (e) correspond to silicon, carbon, and hollow sites of SiC, respectively. STEM image showing nanoscale variations in Gr- $InO_2$ interlayer distance due to Gr-O bonding (f). Ground state,

relaxed InO$_2$ structure, calculated via DFT, showing intermittent Gr-O bond formation due to In vacancies (g). 1.9 Å and 2.8 Å in (f) and (g) correspond to interlayer distances between Gr-InO$_2$ with/without Gr-O bonding. Comparison of z-axis STEM intensity profiles of simulated structure in (g) and experimental STEM image in (f) with/without Gr-O bonding, verifying a match for the interlayer distances.

ReaxFF simulations confirm that 2D metals tend to diffuse outward in defined graphene structures when heated to 600°C in an O$_2$ environment. Here, calculations are conducted using Ga metal, due to availability of force field data, to provide baseline information for In behavior at the EG/SiC interface. This is possible because In and Ga belong to the same periodic table group and are chemical analogs. As the number of metal layers increases, de-intercalation becomes more pronounced, resulting in a thinning of the metal layer, particularly near the edges (Figure S9b, e, h). Notably, the model with three metal layers shows the highest de-intercalation rate, leading to a Ga islanding at the center, while only two layers, and eventually one, remain toward the outer edges (Figure S9c, f, i). Density functional theory calculations presented in Figure S10 reinforce this observation by indicating that the strong covalent interactions between the 1st Ga layer and SiC stabilize the Ga layer at the interface, with a binding energy of 13.63 eV/layer, thereby slowing down Ga diffusion from the edges (Figure S10a). However, as additional Ga layers are added, the binding strength between them decreases—the binding energy of the 2$^{nd}$ Ga layer to the 1$^{st}$ is 9.43 eV/layer and the binding energy of the 3$^{rd}$ Ga layer to the 2$^{nd}$ drops significantly to 6.52 eV/layer. Reduction in binding strength increases the mobility of the metal atoms at the interface, directing them towards the edges where high adsorption sites are located. This outward metal diffusion during oxidation leads to the formation of monolayer InO$_2$ with one layer of In and two layers of O as seen in Figure 2a, d, e. The In/O atomic ratio calculated from AES is 0.56 ∓ 0.19, supporting an InO$_2$ stoichiometry.

Monolayer InO$_2$ is thermodynamically preferred at the EG/SiC interface. Cross-sectional STEM (Figure 2d, e) provides a foundation for exploring structural relaxations via DFT for monolayer InO$_2$ (2L oxygen and 1L indium) and bilayer In$_2$O$_3$ (3L oxygen and 2L indium)—initialized at sites projecting onto the silicon (Si), carbon (C), and hollow sites (H) of SiC substrate. The simulated monolayer structure matches experimental observations (Figure 2a, d, e), supporting that this is likely the preferred structure because: (1) the detachment of extra layers during structural relaxations of multilayered initial structures and (2) the chemical stability understood by basic chemistry rules and confirmed by free energy comparison against other candidate structures. When initial structures with more than one layer of indium and two layers of oxygen are relaxed, two types of outcomes are observed: the extra layer of indium and oxygen are detached, leaving two layers of oxygen and one layer of indium attached on the SiC substrate (Figure S11); some indium from the first indium layer is ejected such that the original fully-filled triangular lattice of In becomes 2/3-filled honeycomb lattice. While multilayer $In_2O_3$ can maintain stability over a certain number of layers when fully confined underneath graphene, it is polymorphic. Cross-sectional STEM provides evidence that 6 layers thick $In_2O_3$ (Figure 2b) evolves from being epitaxial to SiC (first 2 layers) to a cubic structure. The interlayer distance is ~2.5 Å, which aligns with the interlayer distance found in cubic $In_2O_3$ (001).[17] This suggests that there is a special relationship of the first few layers to SiC, which is lost for thicker $InO_x$. The preference for monolayer $InO_2$ formation using semi-confined growth scheme with patterned graphene can be explained by charge neutrality and valence shell filling. Indium ions most commonly have -3 formal charge, while oxygen ions without any covalent bond tend to have -2 formal charge (e.g. bulk $In_2O_3$). Oxygen having one covalent bond tend to have -1 formal charge, due to a similar full-shell stability according to the octet rule (e.g. $[OH]^{-1}$). In our system, oxygen has one covalent bond with Si in SiC substrate and a -1 formal charge. Starting from the Si atom on the surface of the substrate, the formal charge on each atom is $\{Si^0 : O^{-1} : In^{-3} : O^{-2} : In^{+3} : O^{-2} : In^{+3} : \dots \}$. Assuming all indium and oxygen atoms are fully commensurate with the substrate (equal number of atoms in each layer for SiC and $InO_x$), the $InO_x$ layering above the substrate can't extend indefinitely due to excess net charge. In fact, monolayer structure with charge balance $\{Si^0 : O^{-1} : In^{-3} : O^{-2}\}$ and equal layer-by-layer stoichiometry, already creates a charge-neutral and chemically stable structure where all shells are filled. This explains the observed detachment of extra layers upon the multi-layer structural relaxation and why, in bilayers that show no detachment, indium atoms are ejected from the first layer to maintain charge neutrality. Hence, the thermodynamic ground state is reached when graphene is patterned and indium outward diffusion is allowed via semi-confined growth scheme, while $InO_x$ is fully confined underneath continuous graphene sheet, leading to the formation of multilayer (0L-6L) $InO_x$.

Density functional theory indicates two nearly energetically degenerate structures exist with Si-H-C and Si-C-H stacking sequences (Table S1), similar to monolayer $GaO_2$.[3] Both of these structures are experimentally observed in the cross-sectional STEM images, where nanoscale variations in the atomic positions of indium and top oxygen atoms are evident in Figure 2d, e. This surface reversing is also observed in the surface oxides of III-nitrides.[18] In both structures, the indium and oxygen atoms occupy distorted edge-shared octahedral and corner-shared tetrahedral positions, respectively, due to the polarity of the EG/SiC interface. Although the structures above hold the basis for the monolayer $InO_2$ structure, the experimentally observed structures are more complex. The distance between the top oxygen in $InO_2$ and graphene exhibits spatial variations ranging from 1.6 Å to 3 Å (Figure 2f), suggesting the formation of Gr-O covalent bonds, similar to C-O-Al bonding observed in graphene grown on sapphire.[19] To investigate this, we created 44 crystal structures by modifying the lowest energy monolayer $InO_2$ structure with Si-H-C stacking sequence (Figure 2d) and relaxed through *ab initio* DFT. We find that Gr-O bonding becomes energetically favorable in the presence of an indium vacancy by 1.57 eV/defect. Specifically, two to three Gr-O bonds are likely to form where there is one indium vacancy, which is explained by charge difference plot (Figure S12). Additionally, clustering of indium vacancies and Gr-O bonds (Figure 2g) is energetically more favorable than a sparse and uniform distribution of indium vacancies by 0.47 eV/defect (Figure S12). This likely results from regions with and without Gr-O bonds that favor shorter and larger interlayer distances between the oxygen and graphene layers, respectively, and clustering the Gr-O bonds reduces the energetic cost associated with these compromising interlayer distances. The simulated (Figure 2g) and experimental (Figure 2f) structures are compared by measuring the STEM intensity profiles along the z-axis. The comparison in Figure 2g confirms that the distances between the top oxygen and indium atoms, both with and without the Gr-O bond, match in the simulated and experimental structures. This supports that the simulated structure (Figure 2g) is very close to the experimental product. The non-van der Waals (vdW) interface between $InO_2$ and graphene is also verified by our graphene exfoliation experiments, in which $InO_2$ is peeled off along with graphene following nickel deposition and exfoliation using Scotch tape.[20] The structural differences between the graphene-intercalated 2D $InO_x$ structures in this study and those described in Ref[8], i.e. monolayer thickness and Gr-O bonding, are likely due to variations in the synthesis methods employed. Specifically, our study employs patterned graphene, in contrast to the continuous graphene used in Ref[8], which is associated with variations in $InO_x$ thickness, typically resulting in the formation of multilayer $InO_x$.

The monolayer $InO_2$ exhibits a direct band gap of ~4.1 eV. Using the monolayer $InO_2$ structure in Table S1 (Si-H-C stacking), the electronic band structure of the $InO_2$/SiC heterostructure was calculated via DFT. Initially, we validated our DFT approach by calculating the band structure of pristine 4H-SiC and 6H-SiC, finding an indirect band gap of 2.2 eV and 2 eV (Figure S13). This value is lower by ~1 eV compared to established data for 4H and 6H-SiC due to known underestimations in DFT calculations.[21] To adjust for this, we added ~1 eV to the calculated band structure results, leading to a calculated direct band gap of ~4.1 eV at the Γ point (**Figure 3a**) for $InO_2$, matching that of earlier theoretical calculations of monolayer InO[22], and larger than indirect band gap observed in bulk $In_2O_3$ (~2.7 eV).[23]

Intercalation of $InO_2$ modifies the absorption characteristics of the heterostructure, while its emission properties remain similar to SiC. The lowest emission energy of the heterostructure matches that of SiC, as both the conduction band (CB) minimum and valence band (VB) maximum are still dominated by SiC states (Figure 3a). Additionally, there is an extra emission peak at 3.5 eV, which arises from transitions between CB minimum of SiC and VB maximum of $InO_2$. Regarding absorption, prior to the intercalation, the optical gap (the energy gap at $\Delta k = 0$) in 4H-SiC is 4.4 eV at the M point.[24] With $InO_2$ intercalation, new states at the CB emerge at the Γ point, reducing the smallest direct bandgap of the heterostructure to 3.72 eV. Low-loss EELS measurements, which correlate to the joint density of states (JDoS), were employed to confirm these findings (Figure 3b). This technique is particularly effective for measuring direct band transitions in the material as inelastic scattering events rapidly decrease with increasing momentum change, making indirect transitions with high momentum change less detectable.[25] The background subtracted $InO_2$ EELS spectrum was fitted with a parabolic curve $(E-E_g)^{0.5}$, reflecting the typical JDoS behavior for direct transitions near the band edge.[26] The absorption, observed at 3.6 eV, aligns with the direct transition from SiC to $InO_2$ bands at the Γ point, as predicted by the calculated band structure (Figure 3a). It's noted that the extracted absorption in EELS highly depends on the chosen energy window for the fit, as variations in this window yielded fit results ranging from 3.43 to 3.74 eV (Figure S14).

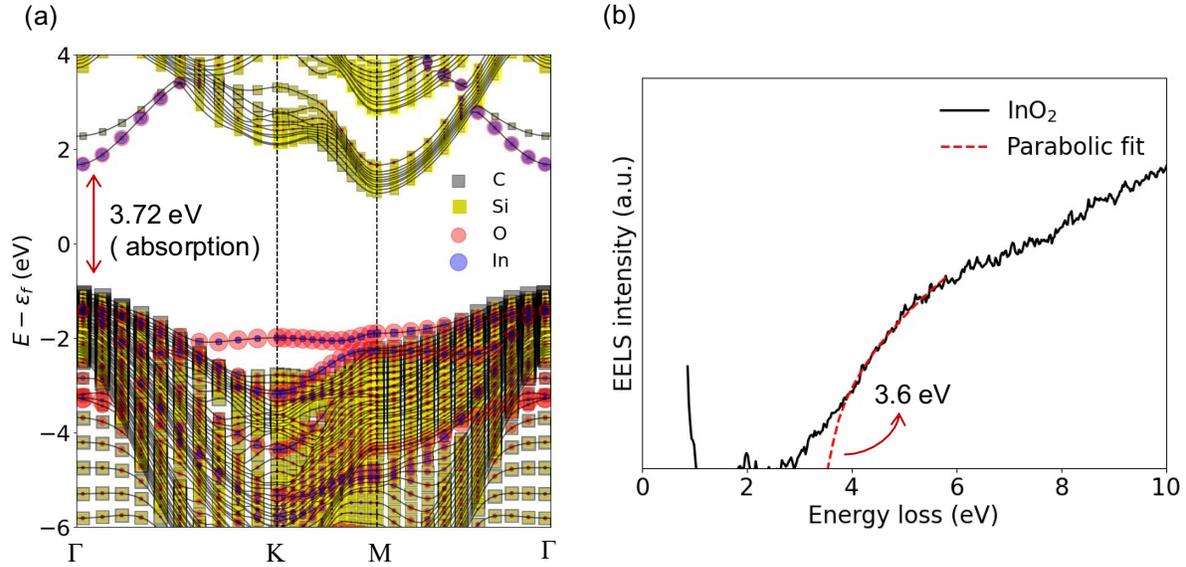

**Figure 3.** Calculated band structure of monolayer $InO_2$ intercalated EG/SiC via DFT, demonstrating a direct band gap of ~4.1 eV for $InO_2$ and 3.72 eV direct transition at Γ point. EELS measurement in STEM acquired from EG/$InO_2$/SiC, demonstrating an absorption at ~3.6 eV, matching with DFT results.

The structural evolution of $InO_2$ from 3D to 2D impacts its phonon band structure. Figure S15a demonstrates the Raman spectra taken from In, $InO_2$ and mixed In/$InO_2$ intercalated EG/SiC. Upon full oxidation, indium ULF peaks (17, 45, and 96 $cm^{-1}$) disappear and $InO_2$ peaks are not detectable. However, in a partially oxidized sample, 2D $InO_2$ peaks emerge at 295, 323, 421, and 450 $cm^{-1}$, likely due to surface enhanced Raman scattering (SERS) effect when near metallic indium. The origin of these peaks is attributed to encapsulated $InO_2$, confirmed by DFT calculations that match with experimental peak positions (Figure S16) Compared to the bulk $In_2O_3$ Raman peaks at 303, 360, and 492 cm-1, [27] 2D $InO_2$ peaks are red shifted which is attributed to the expansion of In-O bonds influenced by Si-O and C-O bonds present. Structural calculations show that the bond distance between the top oxygen and indium increases from 2.05 Å to 2.2 Å due to Gr-O bonding (Figure S18). Likewise, the presence of Si-O bonds increases the bottom O-In distance to 2.25 Å, larger than the one in bulk $In_2O_3$ (2.14 Å),[28] explaining the observed red shift in the Raman spectra due to bond expansion at the EG/SiC interface.

Electronic transport properties of In intercalated EG/n-SiC Schottky diodes were evaluated before and after oxidation, demonstrating a $10^5$ fold enhancement in the rectification ratio as a result of the oxidation. As intercalation transforms EG/n-SiC junction from ohmic to Schottky due to the elimination of buffer layer and $E_F$ depinning,[10] indium intercalated EG/n-SiC diode is also fabricated for comparison. As top and bottom electrodes, 5/40 nm thick Ti/Au is deposited on EG and n-SiC, respectively. Device fabrication steps are shown in Figure S19. First, we assessed the contact properties of the EG/Ti/Au junction (top electrode) to Ohmic behavior and that the barrier observed in diode measurements is originating from the modulations at the EG/n-SiC interface. Using the circular transfer length method (CTLM), the contact resistivity ($\rho_c$) is extracted as $2.06 \times 10^{-6}$ $\Omega cm^2$ (Figure S20), which confirms a low resistance ohmic contact. For the bottom electrode, graphene grown on the C face of SiC facilitates ohmic contact formation with Ti/Au (Figure S21).[10] Indium and InO$_2$ intercalated EG/n-SiC diodes are evaluated by grounding the n-SiC substrate (**Figure 4a)**. Schottky like behavior is verified for both samples, which is attributed to the $E_F$ depinning due to Si-In (In) or Si-O (InO$_2$) bond formation via intercalation, similar to H intercalated EG/n-SiC diodes.[10] However, there is up to a $10^5\times$ difference between the rectification ratio (RR) of In and InO$_2$ intercalated devices with a max RR of ~70 and ~$2\times10^6$, respectively, indicating suppressed reverse leakage current with InO$_2$ intercalation. We note that additional material development is needed for improved device reliability and uniformity. This is evident when comparing device-to-device performance: forward bias $V_{th}$ varies between 0.07 V and 0.43 V for In (0.34 ∓ 0.19 V), and 0.21 V and 1.33 V for InO$_2$ (0.92 ∓ 0.47 V) devices. Similarly, RR is varied between 1.6 - 70 for In and 135 - $2.18\times10^6$ for InO$_2$ devices. For In intercalated devices, this is likely to be a result of device fabrication on step edges, where no intercalant is found or on non-intercalated terrace regions. On the other hand, for InO$_2$ device, this is attributed to several factors, such as InO$_2$ phase variation, Gr-InO$_2$ interlayer distance, indium vacancies, and deintercalated regions.

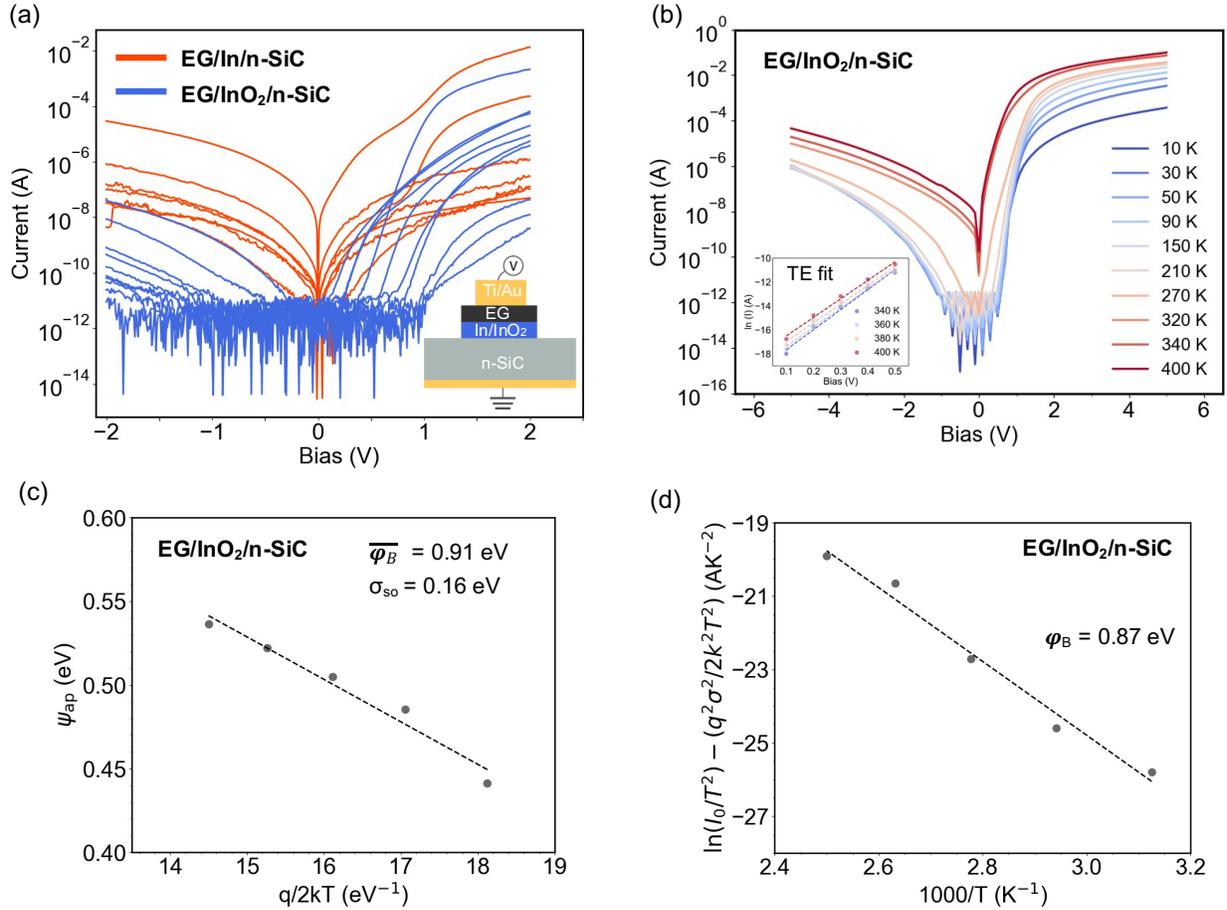

**Figure 4**. Current-voltage (I-V) curves taken from In and InO$_2$ intercalated EG/n-SiC vertical Schottky diodes, demonstrating reduced reverse leakage current with InO$_2$ interfacial layer (a). Schematic for the device structure is given in the inset. Temperature dependent (10 K – 400 K) I-V taken from EG/InO$_2$/n-SiC diode with a forward bias thermionic emission fit given in the inset (b). Apparent barrier height versus q/2kT (c) and modified Richardson plot (d) of the EG/InO$_2$/n-SiC diode according to the Gaussian distribution of the barrier heights.

Temperature-dependent electrical analysis on the EG/InO$_2$/n-SiC junction, along with nanoscale transport properties examined via C-AFM, reveals nanoscale inhomogeneities in the barrier height. We first conducted temperature dependent I-V measurements (10 K – 400 K) with EG/InO$_2$/n-SiC to extract the barrier height (Figure 4b). The current in reverse bias remains constant between 10 K – 150 K, indicating that tunneling is dominant < 150 K. The fit of the reverse bias current at 10 K with Fowler Nordheim Tunnelling (FNT) model exhibits a linear relationship (ln ($I/V^2$) vs $1/V$), indicating tunneling as dominant conduction mechanism for low temperatures (Figure S22a). In the forward bias, the I-V data is examined with the thermionic emission (TE) model (see methods).[29] From the Richardson plot (Figure S22b), the barrier height and Richardson constant is extracted as 0.19 eV and 139 Acm$^{-2}$K$^{-2}$, respectively. Although extracted $A^*$ is close to the theoretical Richardson constant of 4H SiC (146 Acm$^{-2}$K$^{-2}$),[30] the barrier height is lower than expected. As an alternative method, we extracted the barrier height from Equation 4 (see methods) by using theoretical $A^*$ for 4H-SiC for each temperature. As shown in Table S4, with higher temperatures, the barrier height increases from 0.39 eV to 0.49 eV and the ideality factor reduces from 3.87 to 1.76. Such a temperature dependent behavior of $\varphi_B$ and $n$ is attributed to the inhomogeneity of the barrier height. At low temperature, electrons without sufficient energy can only surmount patches with lower Schottky barrier. As the temperature increases, more electrons gain sufficient energy to overcome higher barrier, leading to a higher apparent Schottky barrier height. Based on Tung,[31] a negative correlation between the apparent Schottky barrier and the ideality factor indicates the ideality factor decreases with increasing temperature. Lateral inhomogeneities at the junction have also been confirmed by conductive AFM (C-AFM). The atomic resolution current map, shown in Figure S23a (10 nm × 10 nm), verifies the presence of high and low conductivity regions within the sample. Current-voltage curves acquired from different regions on the sample via C-AFM demonstrate variations in $V_{th}$ in the reverse bias from -4 V to -6 V (Figure S23b), verifying the nanoscale inhomogeneities at the EG/InO$_2$/n-SiC junction. As shown in the post-oxidation AES In KLL map (Figure 1h), nanoscale deintercalated regions exist, which contribute to the low $\varphi_B$ regions as EG/n-SiC junction is ohmic due $E_F$ pinning by the Si dangling bonds.

Given these observations, vertical transport has been analyzed using single Gaussian distribution theory. The lateral inhomogeneity can be described by a Gaussian distribution with a mean barrier height ($\overline{\varphi_B}$) and a standard deviation ($\sigma$).[32] According to single Gaussian distribution theory, $\overline{\varphi_B}$ can be expressed by the following model:

$$\varphi_{ap} = \overline{\varphi_B} - \frac{q\sigma^2}{2kT} \qquad \text{Equation (1)}$$

Where $\varphi_{ap}$ and $\overline{\varphi_B}$ are apparent and mean barrier heights, $\sigma$ and is the standard deviation of the barrier height distribution which is a measure of the barrier height homogeneity. The experimental $\varphi_{ap}$ vs $q/2kT$ is shown in Figure 4c, where a linear relationship is seen. From the slope and the intercept of the $\varphi_{ap}$ vs $q/2kT$ plot, $\overline{\varphi_B}$ and $\sigma$ were extracted as 0.91 eV and 0.156 eV, respectively. Considering the lateral inhomogeneities in the barrier, a modified Richardson model can be expressed as:[30]

$$\ln\frac{J_0}{T^2} - \left(\frac{q^2\sigma^2}{2k^2T^2}\right) = \ln(A^*) - \frac{q\varphi_B}{kT} \qquad \text{Equation (2)}$$

$\ln \frac{J_0}{T^2} - (\frac{q^2 \sigma^2}{2k^2 T^2})$ vs *1/T* should give a straight line with the slope proportional to $\varphi_B$. Figure 4d presents the plot calculated with $\sigma$ obtained from Figure 4c. Linear fit to the plot represents the true activation energy plot, where $\varphi_B$ from graphene to SiC is extracted as 0.87 eV. These electrical measurements with EG/InO$_2$/n-SiC verify Schottky barrier formation with InO$_2$ intercalation and demonstrate the potential of this structure to be used as tunnel or filter barriers in 2D/3D hybrid hot electron transistors (HETs). When utilized as a tunnel barrier, the emitter current density reaches 10$^5$ A/cm², which is significantly higher than those used in graphene-based HETs,[33–35] attributed to the forward-biased MOS-based Schottky diode with a monolayer-thick interfacial layer. On the other hand, when the EG/InO$_2$/n-SiC heterostructure serves as a filter barrier, it can suppress leakage current due to its high rectification ratio (RR). Unfortunately, device reliability remains a challenge, as evidenced by the large standard deviation in V$_{th}$ of 0.47 V and RR of ~10$^4$. This is likely due to nanoscale non-uniformities, such as variations in the EG-InO$_2$ interlayer distance, the presence of indium vacancies, and deintercalated regions.

**Conclusions**

A new member of the 2D dielectric family, monolayer InO$_2$, synthesized by intercalating epitaxial graphene, is introduced. The influence of graphene patterning and graphene lateral size on the thickness and structural properties of 2D InO$_x$ is shown, where monolayer and multilayer indium oxide is grown using patterned and continuous graphene, respectively. Extensive structural analysis via STEM and DFT indicates monolayer InO$_x$ is generally formed at the EG/SiC interface and is explained based on the octet rule and charge neutrality principles. Additionally, DFT simulations confirmed that Gr-O bonding is stabilized by indium vacancies, a consequence of indium deintercalation during oxidation, highlighting the significant effect of the synthesis route on the structure of intercalated compounds. Moreover, monolayer InO$_2$ exhibits a direct band gap of ~4.1 eV, approximately 1.5× the indirect band gap of bulk In$_2$O$_3$ (~2.7 eV). Finally, Schottky diodes with InO$_2$ and In intercalated EG/n-SiC were fabricated, where a significant reduction in reverse leakage current with the InO$_2$ intercalated device is demonstrated, achieving a maximum rectification ratio of 2x10$^6$ (10$^4$× higher than pre-oxidation) with a barrier height of 0.87 eV. These findings illuminate several facets of materials science including control over intercalant thickness through graphene patterning, the synthesis-structure-property relationship, and vertical electronic transport in intercalated EG/n-SiC structures.

**Experimental Section**

*Epitaxial Graphene Growth:* n-type 4H-SiC wafers (Xiamen Powerway Advanced Material) were first cleaned with acetone, isopropyl alcohol (IPA) and nanostrip via ultrasonication. Then, to remove the surface oxide and polishing scratches from the SiC surface, the wafer was annealed in $H_2$ at 1500 °C for 30 minutes (700 Torr, 10%/90% $H_2$/Ar). Then, Si was sublimated from SiC (0001) surface and epitaxial graphene was formed at 1800 °C, 300 Torr for 30 minutes under Ar flow.

*Indium Oxide Intercalation:* Indium intercalation was first performed using an STF-1200 horizontal tube furnace fitted with a 1-inch outer diameter quartz tube. An alumina crucible was used to hold 10 × 10 $mm^2$ EG/SiC substrates, which were placed with EG on the Si face of SiC facing downwards, toward the inside of the crucible. Then, ~30 mg of In powder (Alfa Aesar, −325 mesh, 99.99%) was placed in the crucible directly beneath the EG/SiC substrate. The crucible with the sample and the respective metal precursor was then loaded into the tube furnace and evacuated to ~3 mTorr. After pressurizing the system to 500 Torr with Ar, the furnace was heated to 800 °C with a ramp rate of 20º $min^{-1}$ under an Ar flow of 50 sccm. Metal evaporation was carried out on EG at 800 °C, 500 Torr for 30 minutes and then the furnace was cooled down to room temperature. Oxidation of indium was conducted by annealing the as-intercalated EG/In/SiC under $O_2$/Ar (10/50 sccm) at 600 °C for 30 min in a rapid thermal annealing (RTA) furnace (OTF-1200X-4-RTP).

*Raman Spectroscopy:* Raman Spectroscopy was performed with a Horiba LabRam Raman system using a laser with 532 nm wavelength and power of ~4 mW. An ultra-low frequency filter was used to filter out the Rayleigh signal. Double sweep spectra were taken with an accumulation time of 30 seconds total using a grating with 300 grooves $mm^{-1}$. Raman mapping was done using the SWIFT ultra-fast imaging technique with varying pixel resolution (typically 0.8×0.8 µm). A Si wafer was used for calibration.

*XPS:* The measurements were carried out with a Physical Electronics Versa Probe III equipped with a monochromatic Al Kα X-ray source (hv = 1,486.7 eV) and a concentric hemispherical analyzer. High-resolution spectra were obtained over an analysis area of 200 × 200 $µm^2$ (30 × 30 $µm^2$ for patterned samples) with a pass energy of 29 eV for C 1s and In 3d regions. O 1s and Si 2p regions were acquired with a pass energy of 55 eV. As the samples were electrically contacted to the XPS stage, energy calibration was not done. In this case, Fermi level of the XPS spectrometer aligns with that of sample. This was confirmed by checking the peak positions in the spectra taken with neutralizer on and off.

*Auger Electron Spectroscopy (AES):* AES maps were acquired with Physical Electronics Versa Probe III equipped with Auger Spectroscopy using an electron beam with an energy of 10 keV and a current of 5 nA. For AES mapping, a three-point acquisition method was used for the intensity calculation at each pixel, where a single point is used to define the peak intensity, and two points were chosen to define the background intensity. Maps are the average of 15 and 5 frames for In and O, respectively.

*AFM*: The topography images were acquired with a commercial instrument (Bruker, Dimension Icon) in the PeakForce tapping mode by using a ScanAsyst-Air probe with a constant force of 1 nN.

*C-AFM:* The C-AFM experiments were performed under uncontrolled ambient conditions (temperatures of 30-35°C, measured near the sample, and relative humidity levels of 30-40%), using a commercial instrument (Asylum Research, Cypher VRS). The experiments utilized a diamond-coated, conductive AFM probe (Nanosensors, CDT-NCHR) with a normal spring constant of 58 N/m. The C-AFM image in Figure S23a ($10 \times 10$ nm$^2$) was recorded with a high scan frequency of 15.62 Hz, under the application of a low bias voltage of 5 mV, conditions conducive to atomic-resolution imaging under ambient conditions.[36] The interaction force between the tip and the sample was kept at the snap-in level, with no additional application of normal force. Current-voltage curves (S23b) were acquired with a commercial instrument (Bruker, Dimension Icon) using an SCM-PIT-V2 probe.

*STEM, EDS and EELS:* For the STEM images shown in Figure 2a, d, e, f, the cross-sectional samples were prepared using a FEI Helios 660 focused ion beam (FIB) system. A thick protective amorphous carbon layer was deposited over the region of interest then Ga+ ions (30kV then stepped down to 1kV to avoid ion beam damage to the sample surface) were used in the FIB to make the samples electron transparent for STEM images. High resolution STEM was performed at 300 kV on a dual spherical aberration-corrected FEI Titan G2 60-300 S/TEM. All the STEM images were collected by using a high angle annular dark field (HAADF) detector with a collection angle of 50-100 mrad. Energy-dispersive X-ray spectroscopic (EDS) elemental maps of the sample surface were collected by using a SuperX EDS system under STEM mode which has four detectors surrounding the sample. The cross-sectional samples in STEM images in Figure 2b, S2, and S3 were prepared using a Focused Ion Beam Scanning Electron Microscope (FIB-SEM) equipped with a Ga ion gun (Thermo Fisher Helios 5). The initial tungsten (W) deposition was performed using electron beam deposition to form a 100 nm protective layer on the surface. This was followed by ion beam deposition of a thicker W film (5–6 μm) over the region of interest to protect the surface from ion damage during FIB milling. STEM micrographs were captured, and elemental maps were obtained through STEM-EDS using a double-corrected HRTEM/STEM (TFS Spectra Ultra, operating at 300 keV) and an Ultra-X EDS detector. The EDS data was processed using Velox 3.0 EDS Software.

For absorption analysis on $InO_2$ intercalated EG/SiC a line scan using EELS under STEM mode was performed across the region of interest. A GIF Quantum 963 system was used to collect all the EELS data. By using a X-FEG high brightness electron gun with a monochromator, the energy resolution is ~0.1eV. EELS spectra were collected using a C2 aperture size of 70 μm, camera length of 38 mm, entrance aperture of 2.5 mm, and a dispersion of 0.1 eV/pixel. This corresponds to convergence and collection semi-angles of 9.3 and 18.7 mrad, respectively. Ideally, graphene should be removed from the $InO_2$ surface to avoid interference in the EELS spectrum from graphene's plasmonic contributions. However, attempts to remove graphene resulted in the detachment of $InO_2$ as well because of covalent bonding between graphene and oxygen at the interface. Consequently, EELS measurements were performed with graphene intact. Initially, the zero-loss peak was subtracted using a power law fit, and then the spectrum from unintercalated graphene was subtracted from that of $InO_2$.

*Device Fabrication:* The Schottky diodes were fabricated via optical and e-beam lithography. The steps before $InO_2$ formation, i.e. alignment mark and graphene etching, were made with optical lithography, while the steps after were made with e-beam lithography. Device fabrication steps are shown in Figure S19. First, alignment mark layer was made by etching n-SiC ~500 nm deep with $SF_6$ plasma. Then, circular graphene patterns with diameter ranging from 3 μm to 30 μm were etched with $N_2$ plasma in 15 seconds. Following the $InO_2$ formation, graphene was contacted via Ti/Au (5/40 nm) lift-off in a circular shape with 1 μm diameter. Note that device active area is ~0.28 μm$^2$ as $Al_2O_3$ etch window is circular with 0.6 μm diameter. The recipes used for the device fabrication are given in Table S3. To measure the contact resistance of graphene/Ti/Au, EG was grown on insulating 6H-SiC (0001). The details of the contact resistance measurements and calculations are given in Figure S20. After contact lift-off, graphene was etched by $N_2$ plasma outside of active area to minimize vertical transport through non-intercalated graphene patches. In this layer, no lithography is used as metal contact protects underlying $InO_2$/EG from plasma damage. Since as-grown EG/n-SiC junction is ohmic, this etching step is crucial to minimize regions with low Schottky barrier height (SBH) and obtain higher rectification ratio (RR). Before depositing $Al_2O_3$ isolation layer, a seed layer of Al (2 nm) was deposited on the sample using e-beam evaporation (0.1 Å/sec). Then, 30 nm $Al_2O_3$ was deposited via ALD at 150 °C using trimethyl aluminum and water. This layer is to prevent leak from metal leads to the n-SiC substrate. To open the Ti/Au contacts, $Al_2O_3$ was etched using $BCl_3/Cl_2$ plasma for 60 seconds via e-beam lithography, corresponding to 35 nm $Al_2O_3$. Over etching was employed to make sure all the $Al_2O_3$ is cleaned. Since Au is a good etch stop for $BCl_3$ and $Cl_2$ plasma, the sample easily survives with longer etching. For the metal leads and pads, Ti/Au (10/120 nm) lift-off was employed using e-beam lithography.

*Models for Electrical Transport Analysis*: In the forward bias, since current increases with temperature, the I-V data is examined with the simplified thermionic emission (TE) model,[29] as described below:

$J = J_0 \times \exp(\frac{qV}{nkT})$  Equation (3)

$J_0 = A^* T^2 \times \exp(\frac{-q\varphi_B}{kT})$  Equation (4)

where $J$ is the forward bias current density, $J_0$ is the saturation current density, $V$ is the applied bias, $n$ is ideality factor, $k$ is Boltzmann constant, $T$ is temperature, $A^*$ is the Richardson constant of 4H-SiC, and $\varphi_B$ is the barrier height from EG to SiC. Equation (3) can be rewritten as:

$\ln(J) = \ln(J_0) + \frac{qV}{nkT}$  Equation (5)

where *ln (J)* vs *V* plot should yield linear relationship if thermionic emission is the dominant mechanism in the forward bias. At low bias (<0.5 V), *ln (J)* vs *V* plot yields a straight line (Figure 4b inset) for T > 320 K. Note that linearity in *ln (J)* vs *V* is lost for V > 0.5 V, which is attributed to the activation of FNT at higher bias. To extract the barrier height, $\varphi_B$, from forward bias characteristics, saturation current density, $J_0$, was extracted at temperatures 320 K - 400 K from the y-axis intercept of the ln *(J)* vs *V* plot. Then, using the Equation 6, which is modified from Equation 4, barrier height ($\varphi_B$) and Richardson constant ($A^*$) are extracted from the slope and intercept of $\ln \frac{J_0}{T^2}$ vs *1/T* plot, as 0.19 eV and 139 Acm$^{-2}$K$^{-2}$, respectively (Figure S19b).

$$\ln \frac{J_0}{T^2} = A^* \times \frac{-q\varphi_B}{kT} \qquad \text{Equation (6)}$$

*DFT:* Structural stability calculations for InO$_x$ were done by ab-initio DFT implemented by software package, VASP. Projected Augmented Wave method with PBE functional and plane waves basis set were used. Gamma centered k-mesh were used with step size around 2Pi/24 A-1 along the two dimensions of the slabs. All structures were relaxed with fixed unit cell whose lattice parameters along the two relevant dimensions are set to be those of [0001] of 4H-SiC.

Metal binding energy calculations with different thickness (Figure S9) were conducted using Quantum Espresso. All structural relaxations in this work were performed with pseudopotentials based on the Projector Augmented Wave method. The Perdew−Burke−Ernzerhof (PBE) form of the Generalized Gradient Approximation (GGA) was employed to model the exchange-correlation functional. A 4 × 4 × 1 k-point mesh within a Gamma-centered Monkhorst–Pack scheme was applied for Brillouin Zone integration, with kinetic energy and density cutoffs set at 60 Ry and 600 Ry, respectively, for a supercell with dimensions of 9.23 × 9.23 Å². The Marzari–Vanderbilt cold smearing scheme was used with a broadening parameter of 0.01 Ry. During geometry optimizations, the system was fully relaxed using the Broyden–Fletcher–Goldfarb–Shanno (BFGS) algorithm, with a total energy convergence threshold of 0.0001 Ry and a force threshold of 0.001 Ry/Å. van der Waals interactions were accounted for using Grimme's semiempirical DFT-D3 correction (zero damping). A vacuum layer of 20 Å was introduced perpendicular to the graphene sheets to minimize spurious interactions due to periodic boundary conditions. In the calculations, Si-terminated 6H-SiC (0001) was used, with the bottom layer saturated with hydrogen to reduce computational cost. Gallium atoms were deposited on the top sites of the surface Si atoms, with the second and third layers following an ABC stacking sequence starting from the substrate surface.[1]

For the calculated phonon band structure first-principles calculations based on DFT[37] were performed using the GPAW code,[38,39] which employs the projector-augmented wave (PAW) method for electron–ion interactions.[40] The Kohn–Sham potentials were evaluated on a real-space grid with a spacing of 0.175 Å. The calculations used the linear combination of atomic orbitals (LCAO) mode, expanding Kohn–Sham wavefunctions in the double zeta polarized (dzp) basis set,[41] which reduces computational cost due to its compact size. For the exchange-correlation functional, the Generalized Gradient Approximation (GGA) in the PBE formulation was adopted.[42] The structure was modeled as a slab with a vacuum region of 15 Å in the out-of-plane direction to minimize interactions between periodic images. A Monkhorst–Pack 3×3×1 k-point grid was used for structural relaxation until residual atomic forces were below 1 meV Å$^{-1}$.[43] The optimized lattice parameters were 9.13 Å and 9.14 Å for the a and b lattice constants, respectively. Phonon frequencies and eigenvectors were computed within the finite displacement method (FDM) in the harmonic approximation using Phonopy,[44] interfaced with ASE.[45] A default displacement magnitude of 0.01 Å was used without additional supercell geometry for phonon and Raman calculations. Since DFT operates at zero temperature, temperature effects were not considered. Resonant Raman spectra were computed using our in-house code[46] implemented for GPAW version 22.1, based on third-order perturbation theory.[47] The code requires GPAW to operate in LCAO mode, and post-processes the electron–phonon, and momentum matrix elements into Raman tensors in the finite difference approach. While our implementation relies on GPAW version 22.1, GPAW has since introduced an official Raman code in version 23.6.0.

*ReaxFF Molecular Dynamics Simulations:* Since both In and Ga belongs to the same group in the periodic table and are chemical analogs of each other, a ReaxFF reactive force field developed by Niefind *et al*.[48] for Graphene/Ga/O/SiC interactions was utilized in large scale molecular dynamics simulations. Each model was placed in a box with the cell dimensions of 19.3 nm x 13.7 nm x 1000 nm and each model consists of a Si-terminated 6H-SiC (0001) substrate, a bilayer graphene layer, and Ga layers ranging from 1 to 3 layers. Upon minimization, each model was first kept in an isothermal-isobaric ensemble (NPT) at 300 K to relieve any artificial strain resulting from lattice mismatches at the interface of SiC/Ga/graphene, then subjected to annealing at a target temperature in a constant temperature and constant volume ensemble (NVT) with a temperature damping constant of 100 fs either without or under the $O_2$ exposure. The details of temperature and time durations are given in the captions of the figures (Supporting Information). The time step was set to 0.1 fs. To control the temperature fluctuations, Berendsen thermostat was deployed. Newton equations of motion was integrated using Velocity verlet algorithm. VESTA was utilized for atomic illustration purposes. For the SiC/graphene model where the first layer of bilayer graphene is rotated by 30° with respect to the SiC substrate with the 10 layers to reduce the lattice mismatch between the SiC and graphene to 0.2%. Additionally, to observe the graphene edge effect on the Ga de-intercalation, thus mimicking the experimental setup, the periodicity of bilayer graphene was removed in all three directions

**Acknowledgement**

F.T. would like to thank Fulbright Turkish Commission for the Ph.D. scholarship. Funding for the experimental work done by F. T., D. Z. and Z. J. comes from the Air Force Office of Scientific Research (AFOSR) through contract FA9550-19-1-0295. Electron microscopy work on indium intercalated structures was performed at the Canadian Centre for Electron Microscopy a core research facility at McMaster University (also supported by NSERC (Natural Sciences and Engineering Research Council of Canada) and the Canada Foundation for Innovation. B.P. and N.B. were supported by AFOSR Grant No. FA9550-23-1-0275 for electron microscopy research. C.D. acknowledges financial support from the 2D Crystal Consortium, National Science Foundation Materials Innovation Platform, under cooperative agreement NSF DMR-1539916. G. O. and M. Z. B. acknowledge financial support from the AFOSR Award No. FA9550-22-1-0418. The co-authors acknowledge the Penn State Materials Characterization Lab for use of the Bruker Icon AFM and Physical Electronics Versa Probe III XPS/Auger, and Horiba Raman Spectroscopy tools and Penn State Materials Research Institute staff members Guy Lavelle, Andrew Fitzgerald, Maxwell Wetherington, Jeffrey Shallenberger, and Robert Hengstebeck for fruitful discussions.

**Conflict of Interest**

The authors declare no conflict of interest.

**Data Availability**

The data that support the findings of this study are available from the corresponding author upon reasonable request.

# SUPPLEMENTARY INFORMATION

# Two-dimensional Indium Oxide at the Epitaxial Graphene/SiC Interface: Synthesis, Structure, Properties, and Devices

*Furkan Turker[1, 2], Bohan Xu[3], Chengye Dong[2, 4], Michael Labella III[5], Nadire Nayir[6, 7, 8], Natalya Sheremetyeva[9], Zachary J. Trdinich[1], Duanchen Zhang[1], Gokay Adabasi[10], Bita Pourbahari[11, 12], Wesley E. Auker[5], Ke Wang[5], Mehmet Z. Baykara[10], Vincent Meunier[9], Nabil Bassim[11, 12], Adri C.T. van Duin[1,8,9,13], Vincent H. Crespi[1, 2, 3, 4, 13], Joshua A. Robinson[1, 2, 3, 4, 9, 13]\**

[1]Department of Materials Science and Engineering, The Pennsylvania State University
[2]Center for 2-Dimensional and Layered Materials, The Pennsylvania State University
[3]Department of Physics, The Pennsylvania State University
[4]Two-Dimensional Crystal Consortium, The Pennsylvania State University
[5]Materials Research Institute, The Pennsylvania State University
[6]Paul-Drude-Institute for Solid State Electronics, Leibniz Institute within Forschungsverbund Berlin eV
[7]Department of Physics Engineering, Istanbul Technical University
[8]Department of Mechanical Engineering, The Pennsylvania State University
[9]Department of Engineering Science and Mechanics, The Pennsylvania State University
[10]Department of Mechanical Engineering, University of California Merced
[11]Department of Materials Science and Engineering, McMaster University
[12]Canadian Centre for Electron Microscopy, McMaster University
[13]Department of Chemistry, The Pennsylvania State University

*Corresponding author's email: jrobinson@psu.edu

**Indium Intercalation**

    **Indium Intercalation with Continuous Graphene**

As epitaxial graphene is high quality, low power $O_2$ plasma treatment is applied to generate defects in graphene (1x1 cm$^2$) prior to intercalation. Then, indium is intercalated by evaporating metallic indium precursor at 800 °C, 500 Torr, under 50 sccm Ar flow in 30 minutes. High resolution X-Ray Photoelectron Spectra are given in Figure S1. SiC peak shift from 283.7 eV to 282.7 eV and the elimination of buffer layer peaks in C 1s show successful indium intercalation.[49] Minimal oxygen peak in O1s, as well as Indium peak position at 443.8 eV verify that graphene protects intercalated indium from oxidation, similar to the Ref.[1]

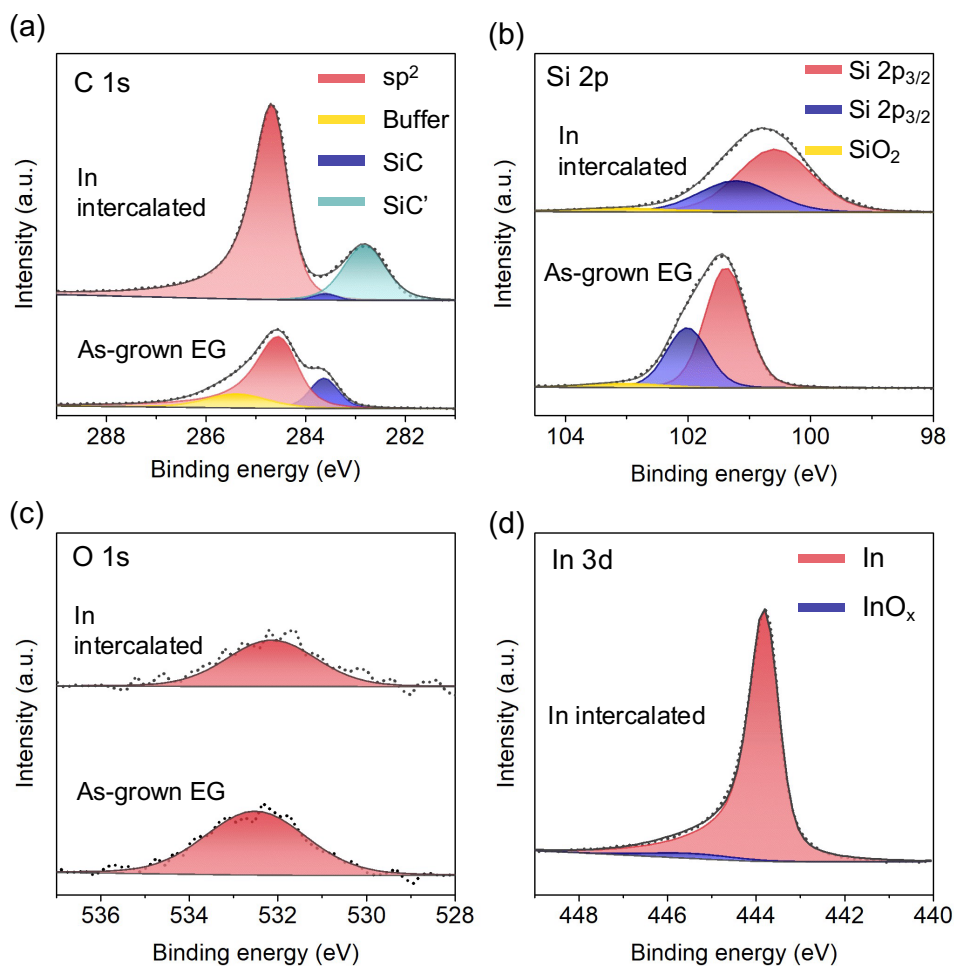

**Figure S1.** High resolution X-ray photoelectron spectra of the continuous EG/SiC sample before and after indium intercalation showing (a) C 1s, (b) Si 2p, (c) O 1s, (d) In 3d regions.

## Indium Intercalation with Patterned Graphene

In the 2nd approach, epitaxial graphene is first patterned by optical lithography and then etched with $O_2$ plasma to create circular graphene with 20 μm diameter. Then, EG/SiC is intercalated with indium via same method as the 1st approach (see methods). Scanning electron microscope (SEM) image of the intercalated patterned graphene is given in Figure S2a. Raman spectrum taken from the circles presents metallic indium peaks in low frequency region (17, 45, 96 cm$^{-1}$).[16] Uniformity of the indium intercalation is verified by Raman mapping (Figure S2c, d), scanned for graphene G band at ~1600 cm$^{-1}$ and indium peak at ~17 cm$^{-1}$. Direct evidence for the bilayer indium intercalation is given in the cross-sectional STEM image, along with EDS elemental maps (Figure S2e).

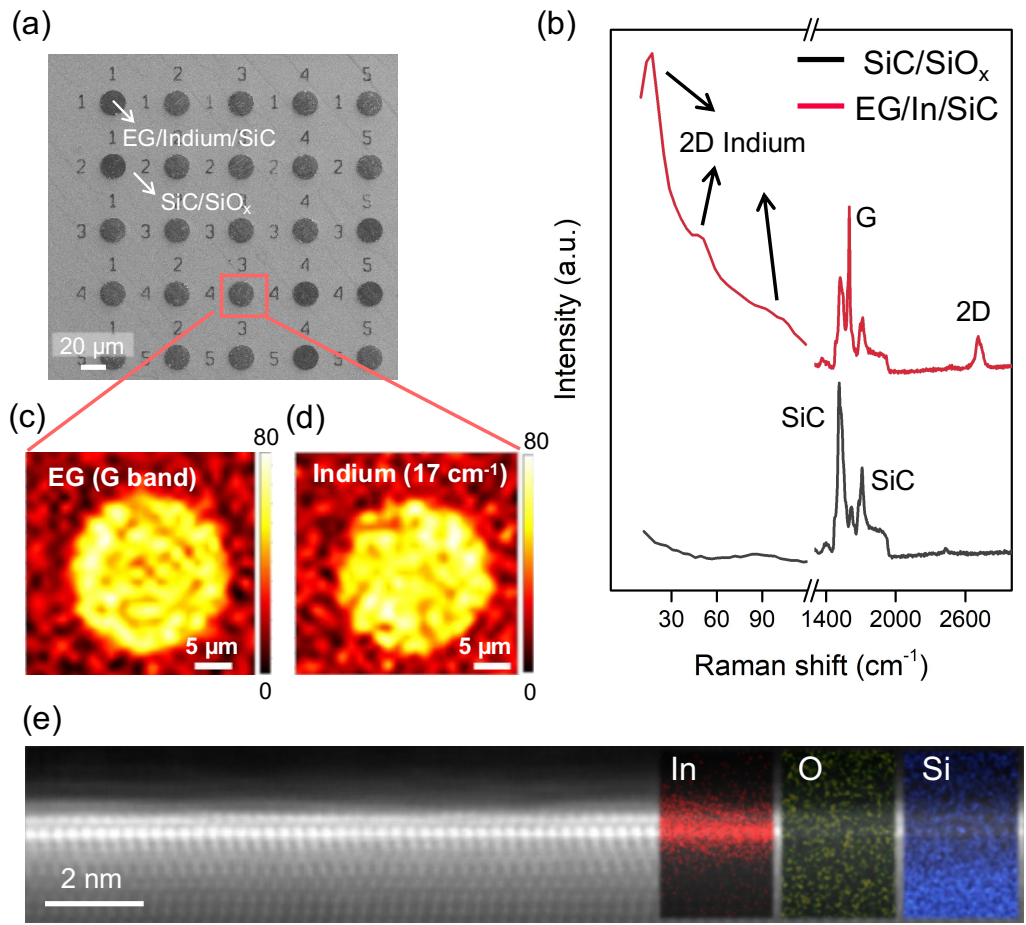

**Figure S2.** Raman, SEM, and STEM characterization of the patterned sample after indium intercalation. SEM image (a) shows that circular graphene patterns with 20 μm diameter are formed. Raman spectrum taken from within the circle exhibits metallic indium ultra-low frequency peaks at 17, 45, 96 cm$^{-1}$. Uniformity of the intercalation is confirmed by Raman mapping where graphene G band at 1600 cm$^{-1}$ (c) and metallic indium peak at 17 cm$^{-1}$ (d) is continuous within the circle. Direct evidence for the intercalation is given in cross-sectional STEM image along with EDS elemental maps (e), showing epitaxial, bilayer indium at the graphene/SiC interface.

Low power $O_2$ plasma treatment for graphene defect generation is not necessary for successful intercalation using the patterned graphene sample. Following graphene etching during lithography, the edges of the patterns remain defective (Figure S3a) as graphene D peak intensity increases toward the edge of the sample. This allows large area indium intercalation while keeping the graphene pristine. Cross-sectional STEM analysis verifies bilayer indium intercalation between EG and SiC and the absence of indium outside of the patterned region, where graphene is not present. (Figure S3b, c).

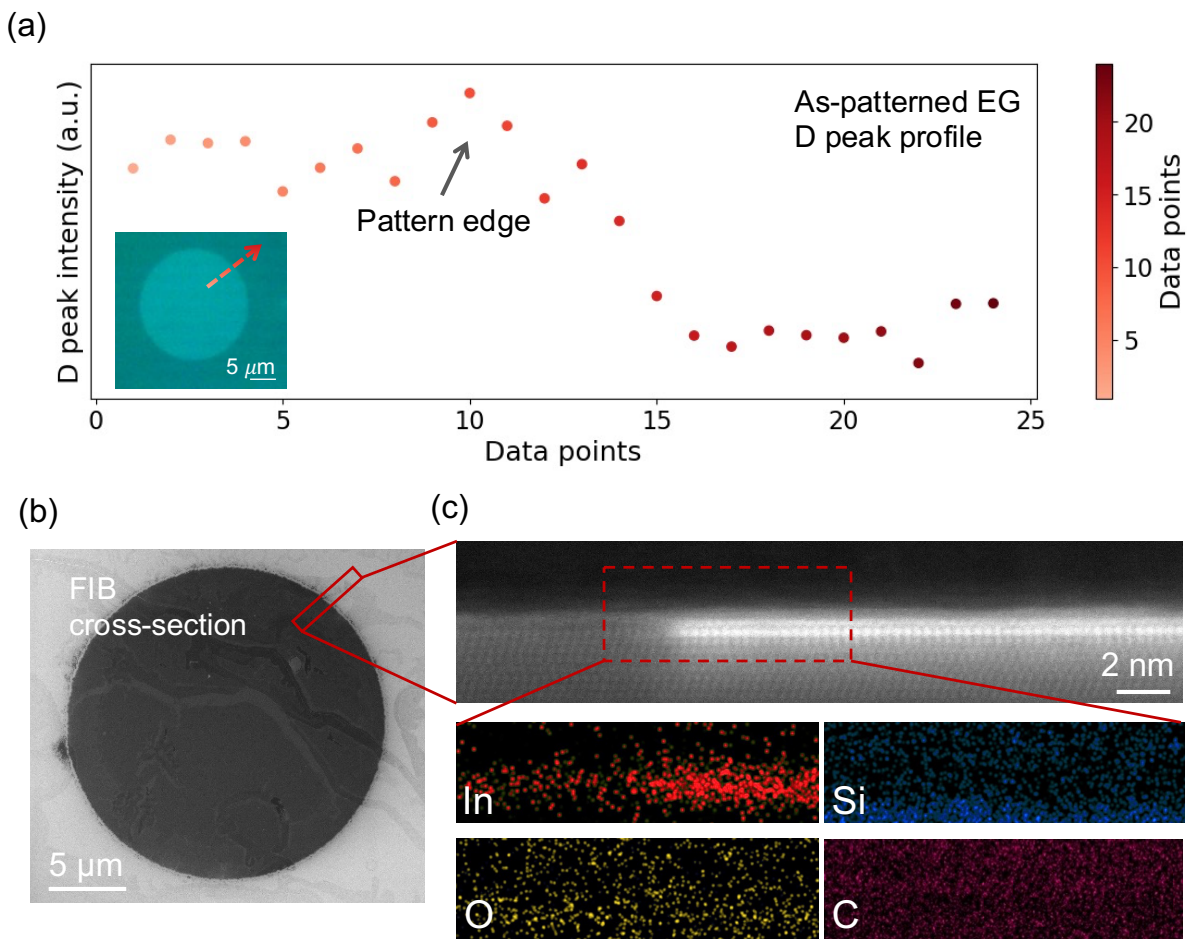

**Figure S3.** D peak intensity of patterned graphene from the center towards the edge before intercalation (a), demonstrating higher defect density at the edge of the pattern due to plasma etching during lithography. SEM image and cross-sectional STEM image along with EDS elemental maps of the patterned sample after indium intercalation, showing that 2D indium film forms underneath graphene.

Following the indium intercalation through patterned EG/SiC, bright features appear in optical micrographs (Figure S4b). Based on Auger Spectroscopy and the observation of graphene wrinkles in AFM near the particle (not shown here), these regions correspond to thick indium particles beneath graphene, which serve as indium sinks during intercalation. During oxidation, graphene initially cracks in these regions, presumably due to high stress (Figure S4d), leading to indium deintercalation. Importantly, the density of these particles reduces by increasing the graphene pattern size and diminish when the intercalation is conducted using the whole sample (1x1 cm$^2$). To reduce the density of these indium particles, graphene is first patterned into large squares (~1.5 mm) and then further etched in the form of circles with 20 µm diameter via another lithography step (2 step etching via lithography). Note that indium intercalation is more successful when graphene is etched via lithography even if the graphene size is 1.5 mm X 1.5 mm than using the whole sample (1 cm X 1 cm), demonstrating indium can diffuse long distances through defective graphene edges (Figure S5). Optical micrograph and Raman indium peak mapping after 2$^{nd}$ etching step are given in Figure S4f, g, verifying smoother surface and uniform intercalation. Importantly, graphene crack formation and indium deintercalation during oxidation are significantly suppressed via this method as SEM image presents smoother surface (Figure S4h).

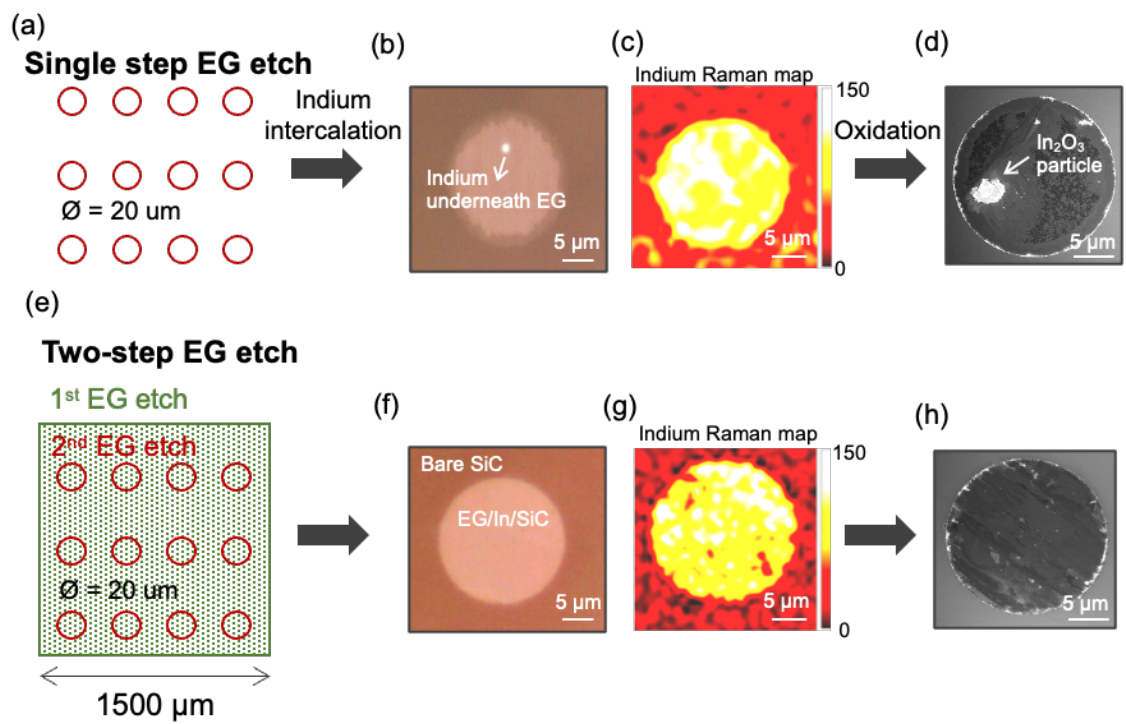

**Figure S4.** Comparison of single and two step etching of graphene for the 2D InO$_x$ formation. Schematics showing the single step (a) and two-step (e) etching of EG where graphene is etched outside of circular patterns. In (e), first, graphene is patterned with 1.5 mm X 1.5 mm size for the indium intercalation and then etched again in the form of circles for the subsequent oxidation. Optical micrographs (b, f) and metallic indium Raman (17 cm$^{-1}$) maps (c, g) taken from In/EG, showing that particle formation beneath EG can be avoided via 2 step etching. SEM images of the oxidized samples for single step (d) and two-step graphene etching (h), showing that graphene crack due to indium particle oxidation underneath EG is avoided via two-step etching.

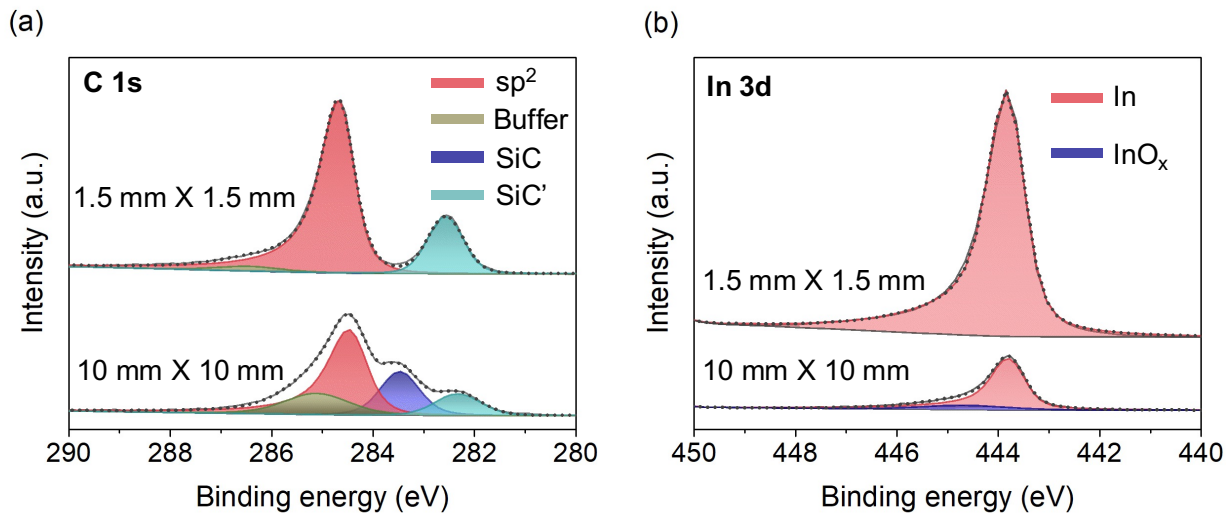

**Figure S5.** XPS high resolution C 1s (a) and In 3d (b) spectra taken from indium intercalated samples with sizes 1.5 mm X 1.5 mm (lithographically etched) and 10 mm X 10 mm (whole sample). SiC peak in C 1s spectra shifts to lower binding energy following indium intercalation (SiC' peak) due to change in band bending with patterned sample (1.5 mm X 1.5 mm). Partial intercalation with 10 mm X 10 mm size sample is verified by the remaining unshifted SiC peak in C 1s spectrum along with lower intensity indium peak in In 3d region.

**InO$_x$ Formation at the EG/SiC Interface**

Oxidation of indium intercalated continuous and patterned graphene yields partial and fully oxidized indium, respectively. Partial oxidation of indium in continuous graphene sample is verified by Auger spectra taken from the bright/dark contrast in the optical micrograph (a). Indium MNN Auger peaks in bright/dark regions at ~403/400 eV correspond to metallic/oxide indium peak positions, respectively. Additionally, O KLL peak intensity in bright region is minimal (Figure S6c) compared to dark contrast and Raman spectrum taken from the bright region still presents metallic indium Raman peak at 17 cm$^{-1}$. On the other hand, In MNN peak at 399.8 eV and the absence of metallic indium Raman peak at 17 cm$^{-1}$ from the patterned sample verify successful indium oxidation. Uniformity of indium and oxygen within the circle is presented in Auger In MNN and O KLL maps in Figure 1h, j in the main text.

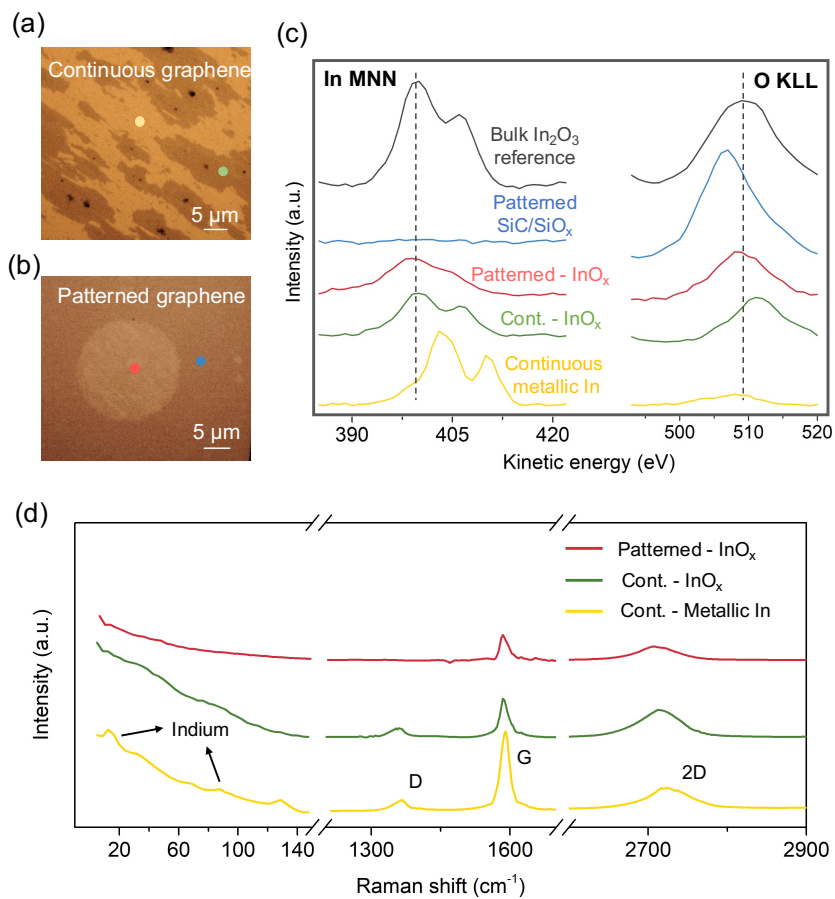

**Figure S6.** Optical micrographs of the indium intercalated continuous (a) and patterned (b) graphene samples after oxidation. Auger spectra (c) from selected regions in these samples, as shown in (a, b). Raman spectroscopy of InO$_x$ intercalated continuous and patterned graphene samples (d). Color code in Auger and Raman spectra correspond to the regions shown in (a, b).

Reduction in the indium thickness to monolayer following the oxidation indicate indium deintercalation when patterned graphene is used. This occurs through the outward lateral diffusion of indium to the edges of the patterned graphene circles, as bulk $In_2O_3$ particles are observed at the perimeter of the circles (Figure S7).

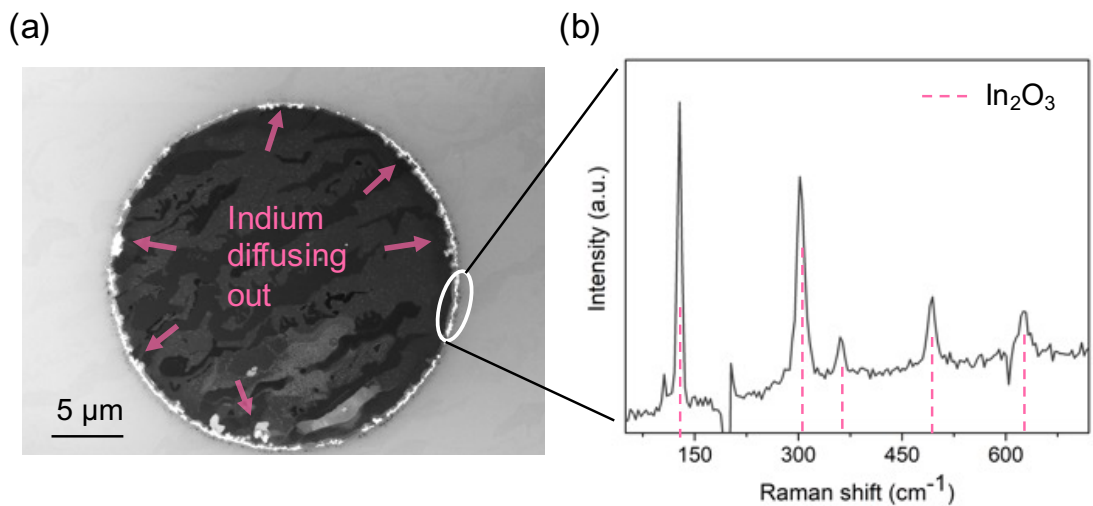

**Figure S7.** Post-oxidation SEM image from EG/$InO_x$/SiC with patterned sample (a), demonstrating lateral outward indium diffusion during oxidation. Raman spectrum taken from the particles at the perimeter of the graphene pattern confirms the $In_2O_3$ particles formation due to indium segregation (b).

Surface root mean square (RMS) roughness ($R_q$) of the oxidized samples grown using continuous and patterned graphene are distinctly different with 1.99 ± 0.25 nm and 0.52 ± 0.02 nm, respectively. AFM image acquired on the continuous sample shows agglomerated $InO_x$ particles on the graphene surface, which is a result of indium deintercalation during oxidation through graphene defects (Figure S8a). Post-oxidation graphene Raman D to G band intensity ratio (I(D)/(IG)) is 0.36 ± 0.06 and 0.07 ± 0.03 for continuous and patterned samples, respectively (Figure S6d). This is not surprising as continuous graphene was treated with low power $O_2$ plasma prior to indium intercalation. Although graphene is healed during intercalation,[1] it does not heal fully as I(D)/(IG) ratio after intercalation is 0.15 ± 0.04. These defects may allow indium deintercalation during oxidation as evidenced from the $InO_x$ particle formation on the graphene surface (Figure S8a). On the other hand, in patterned sample, graphene remains high quality pre-oxidation as plasma treatment is not applied for intercalation. Although indium deintercalates through patterned graphene/SiC interface as well, this mostly occurs laterally to the edges of the patterns, as it will be discussed later.

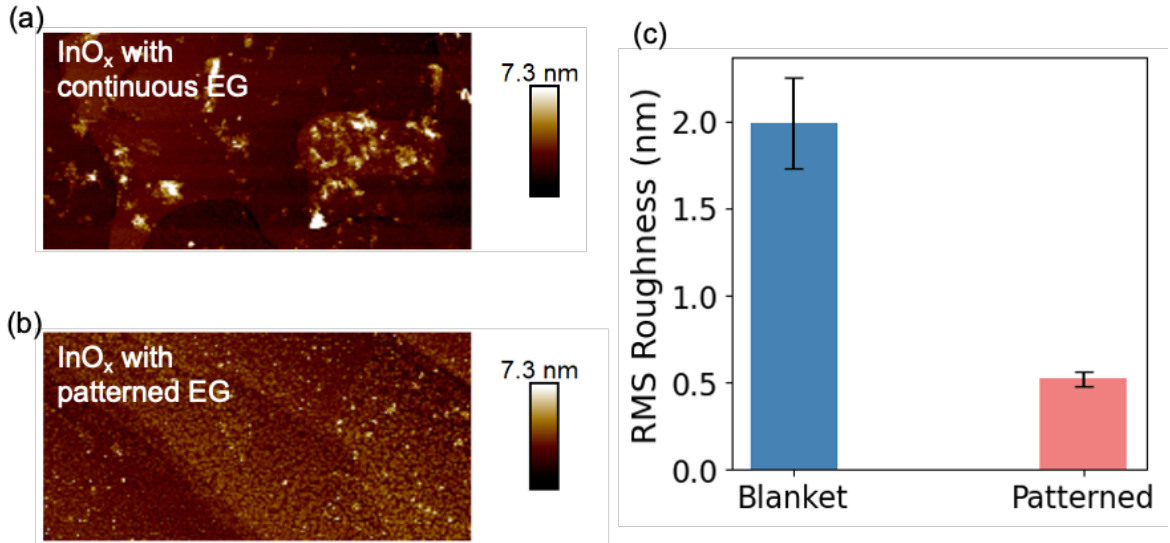

**Figure S8:** Atomic force microscope images of the $InO_x$ intercalated continuous (a) and patterned (b) graphene samples. Root mean square roughness of the graphene surfaces is given in (c), verifying smoother surface with patterned sample due to suppressed indium deintercalation through graphene defects, as well as lateral outward diffusion of In to the perimeter of the circular pattern.

**Molecular Dynamics Simulations via ReaxFF**

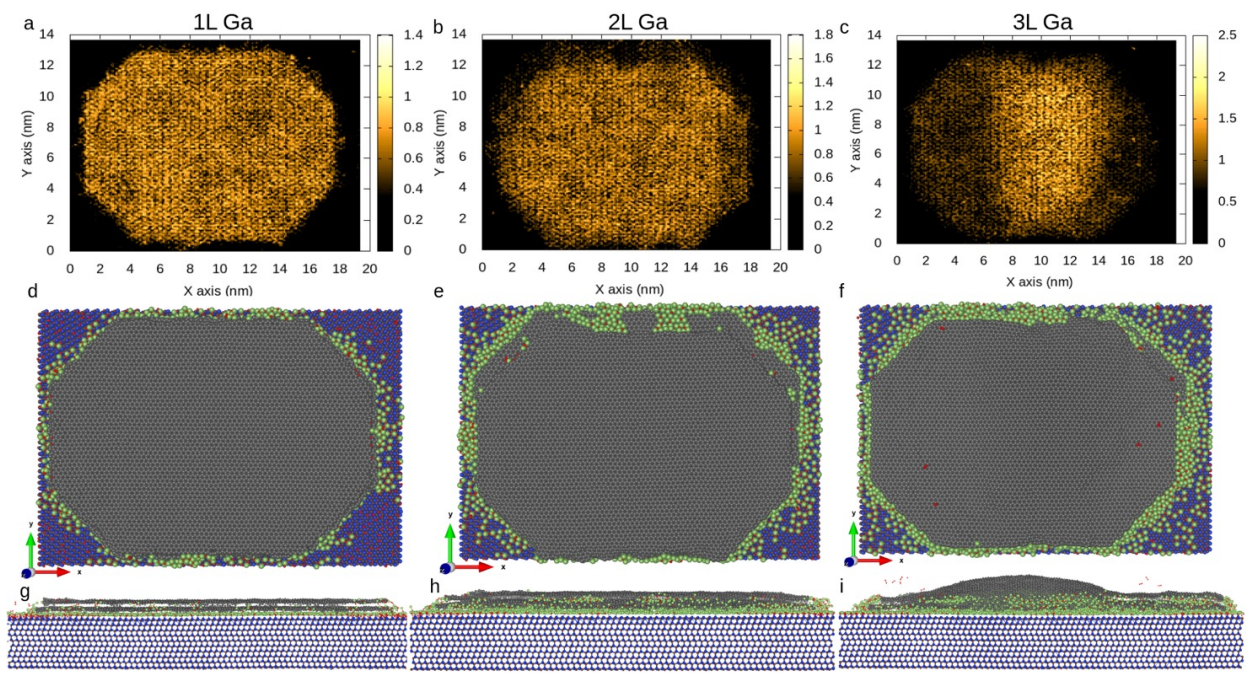

**Figure S9.** Heat maps of the SiC/bilayer graphene models with (a) one Ga (b) two Ga and (c) three Ga layers annealed at 1500K for 150 ps. The brighter regions represent the positions of Ga within the models. Top and side views of the models (d, g) in a, (e, h) in b and (f, i) in b, respectively. Si, C, Ga and O are represented by blue, gray, green and red balls, respectively.

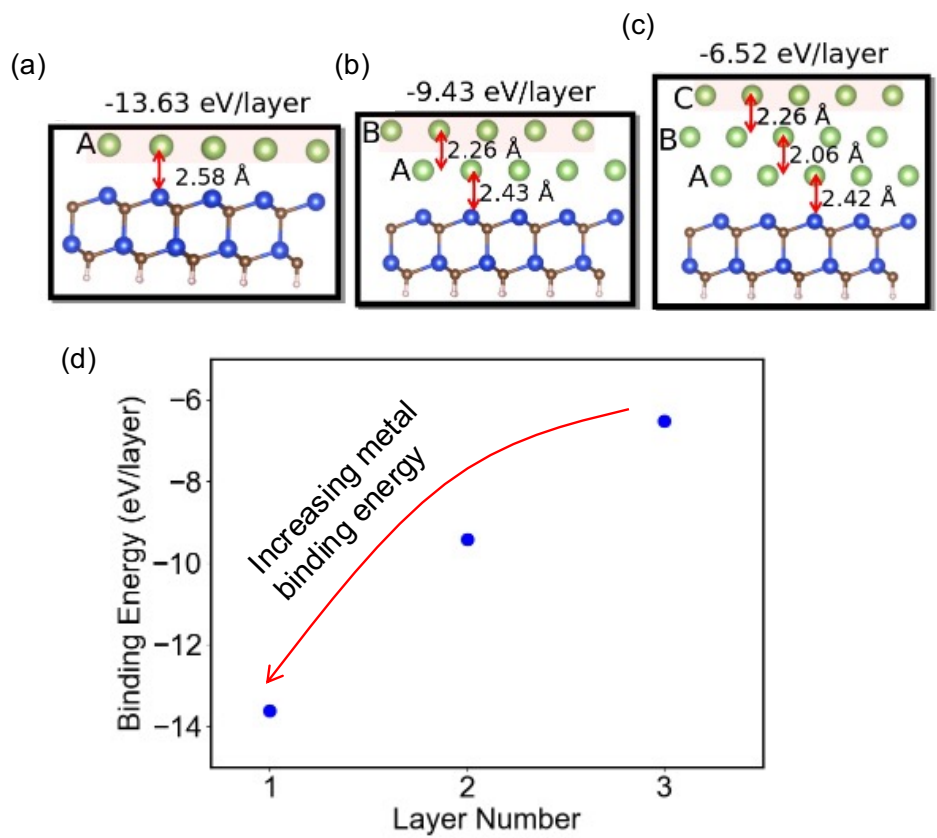

**Figure S10.** Atomic illustrations of the models used in DFT calculations. Side and top views of the models with (a) one, (b) two and (c) three Ga layers deposited on a Si-terminated 6H-SiC(0001) with bottom saturated by H atoms. Ga atoms on the first layer in a deposited on the top site of surface Si atoms, with the second in (b) and third layers in (c) following an ABC stacking sequence starting from the substrate surface. Binding energy vs Ga layer number plot (d). The binding energies of a Ga layer with the SiC surface and other Ga-layers are displayed at the top of each subfigure. Si, C, Ga and O are represented by blue, brown, gray and red balls, respectively.

## Structure of 2D InO₂

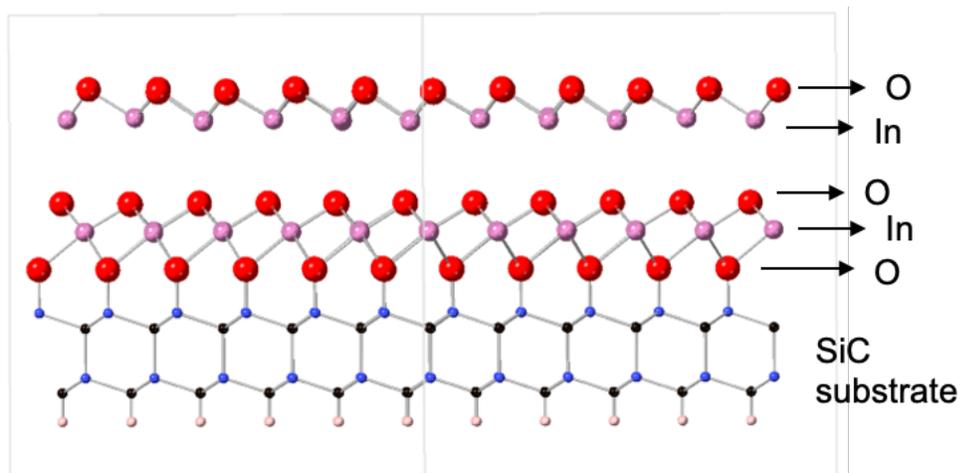

**Figure S11:** Final relaxed multilayered structure often shows detachment of extra layers, even if all the layers are originally in proximity with each other and the substrate in the initial structure prior to structural relaxation. This suggests the chemical stability of the monolayer structures relative to multilayer structures.

**Table S1.** Structural stability calculations via DFT, with indium and oxygen atomic positions projected onto silicon (Si), hollow (H), and carbon (C) sites of SiC. Calculations yield nearly energetically degenerate two structures with Si-H-C and Si-C-H stacking sequence. In atomic structures, Si is represented by blue, C by brown, O by yellow, and In by red colors.

| Bottom Oxygen Position | Indium Position | Bottom Oxygen Position | Relative Stability (eV/indium) | SiC [11-20] view |
|---|---|---|---|---|
| Silicon | Hollow | Carbon | 0 (Ref) | 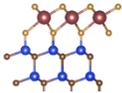 |
| Silicon | Hollow | Silicon | +0.36 | 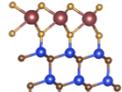 |
| Silicon | Carbon | Silicon | +0.37 | 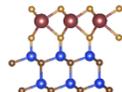 |
| Silicon | Carbon | Hollow | -0.01 | 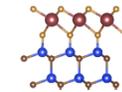 |

To verify Gr-O bonding, we created 44 crystal structures using a 6x6 supercell of the SiC slab, intercalant, and graphene, by modifying the lowest energy monolayer $InO_2$ structure with Si-H-C stacking sequence. These structures include $InO_2$ with varying concentrations of indium vacancies (2/3, 3/4, 4/36, or 3/36 missing indium atoms), oxygen vacancies, excess oxygen, and no vacancies at all. The formation of Gr-O bonds is anticipated by positioning the graphene sheet close to the top oxygen layer. Additional structures were also simulated for specific purposes, to be explained later. All structures were then relaxed through ab initio DFT as described in the methods section. We found that Gr-O bonding becomes energetically favorable in the presence of an indium vacancy (1.57 eV/defect). Specifically, two to three Gr-O bonds are likely to form where there is one indium vacancy. This can be rationalized by considering the formal charge of the system: when graphene bonds with oxygen, the formal charge of the oxygen increases from 2- to 1-, while the carbon's formal charge remains unchanged in low energy structures by altering its hybridization and reducing the number of double bonds, thereby avoiding the creation of high-energy carbon radicals. As the formal charge of the system increases due to Gr-O bonding, compensation is achieved by either adding excess oxygen or creating indium vacancies. Structures without vacancies start with zero formal change, hence simulation data indicates that the formation of Gr-O bonds is energetically unfavorable, as it would increase the formal charge to positive. The same is true for structures with missing oxygen, which already begin with a positive formal charge prior to forming Gr-O bonds (+2.16 eV/defect compared to those with an indium vacancy). The results for the relaxed structures with Gr-O bonding and either excess oxygen or an indium vacancy are detailed below.

**Excess oxygen:** A structure with excess oxygen starts out with negative formal charge. Such structure with Gr-O bonds is indeed lower in energy than the same structure without Gr-O bonds. However, simulation suggests that excess oxygen would break commensuration and result to disordered structure that is not observed in most part of the sample. This is likely due to the chemical potential cost of having excess oxygen atoms under graphene instead of free oxygen gas.

**Indium vacancy:** A structure with missing indium starts out with 3- formal charge, so it is favorable to increase its formal charge to be closer to zero by forming two to three Gr-O bonds, i.e. Gr-O bonding becomes energetically favorable in the presence of an indium vacancy by 1.57 eV/defect. An indium vacancy creates a -3 formal charge as it has three oxygen atoms as nearest neighbors. In geometric relaxations, two or three Gr-O would spontaneously form to change the formal charge to -1 or 0. The reason why only two Gr-O bonds are formed in some cases, is because of the geometric misalignment between the third oxygen and the closest carbon. In some cases, another reason of not forming the 3rd bond is to avoid creating energetically unfavorable radical/unpaired electron. Additionally, a clustering of indium vacancies and Gr-O bonds is energetically more favorable than a sparse and uniform distribution of indium vacancies by 0.47 eV/defect (Figure S12). This likely stems from the fact that regions with and without Gr-O bonds prefer shorter and larger interlayer distances between the oxygen and graphene layers, respectively. Clustering Gr-O bonds reduces the energetic cost associated with these compromising interlayer distances. Additionally, STEM intensity profiles along the z-axis for the simulated structure with clustered In vacancies and Gr-O bonds (Figure S12c, d), and the experimental STEM images, match each other, as shown in Figure 2h in the main text, demonstrating that the simulated structure in Figure S12c, d could be the experimental product.

Instead of using two or three Gr-O bonds to compensate for the 3- formal charge from missing indium, we also considered, for instance, using one Gr-O bond (1+ formal charge) and one missing oxygen (2+ formal charge). We created more structures with various numbers of Gr-O bonds and missing/escaped oxygen which are placed above graphene sheet as free oxygen gas. However, forming only Gr-O bonds without missing/escaped oxygen is still more favorable by ~2.16 eV/defect. Further entropic contribution of free oxygen gas in typical experimental condition is way smaller than this energy scale. Therefore, we expect the Gr-O bond formation is due to In vacancies.

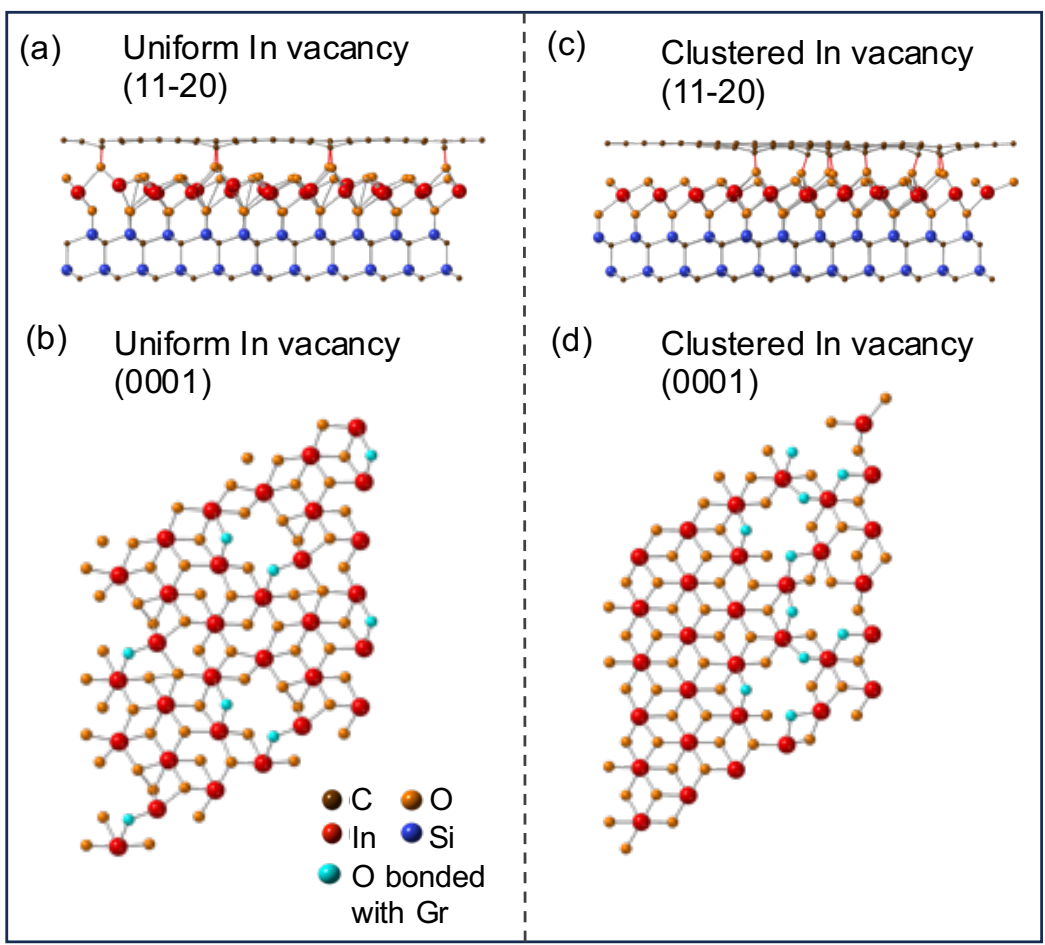

**Figure S12.** Lowest energy monolayer $InO_2$ structures with 4/36 indium vacancies and Gr-O bonding. (11-20) and (0001) views of structures with uniform (a, b) and clustered (c, d) In vacancy and Gr-O bonding distribution. Clustering of indium vacancies is 0.47 eV/defect more energetically favorable than uniformly distributed indium vacancies. O atoms bonded with C in graphene are highlighted in turquoise.

**Electronic and Phonon Band Structure Modifications in EG/InO$_2$/SiC**

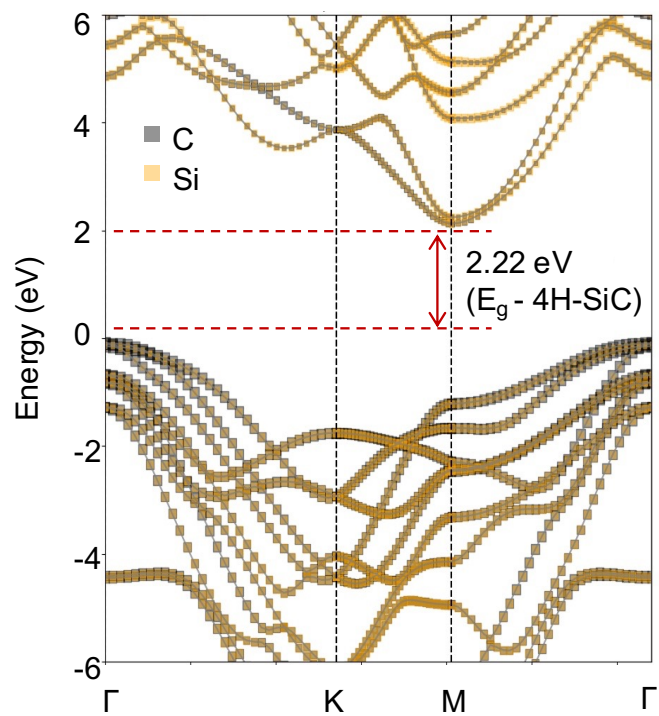

**Figure S13.** Calculated band structure of pristine 4H-SiC, demonstrating underestimated bandgap of 2.2 eV (1 eV lower than the experimental value).

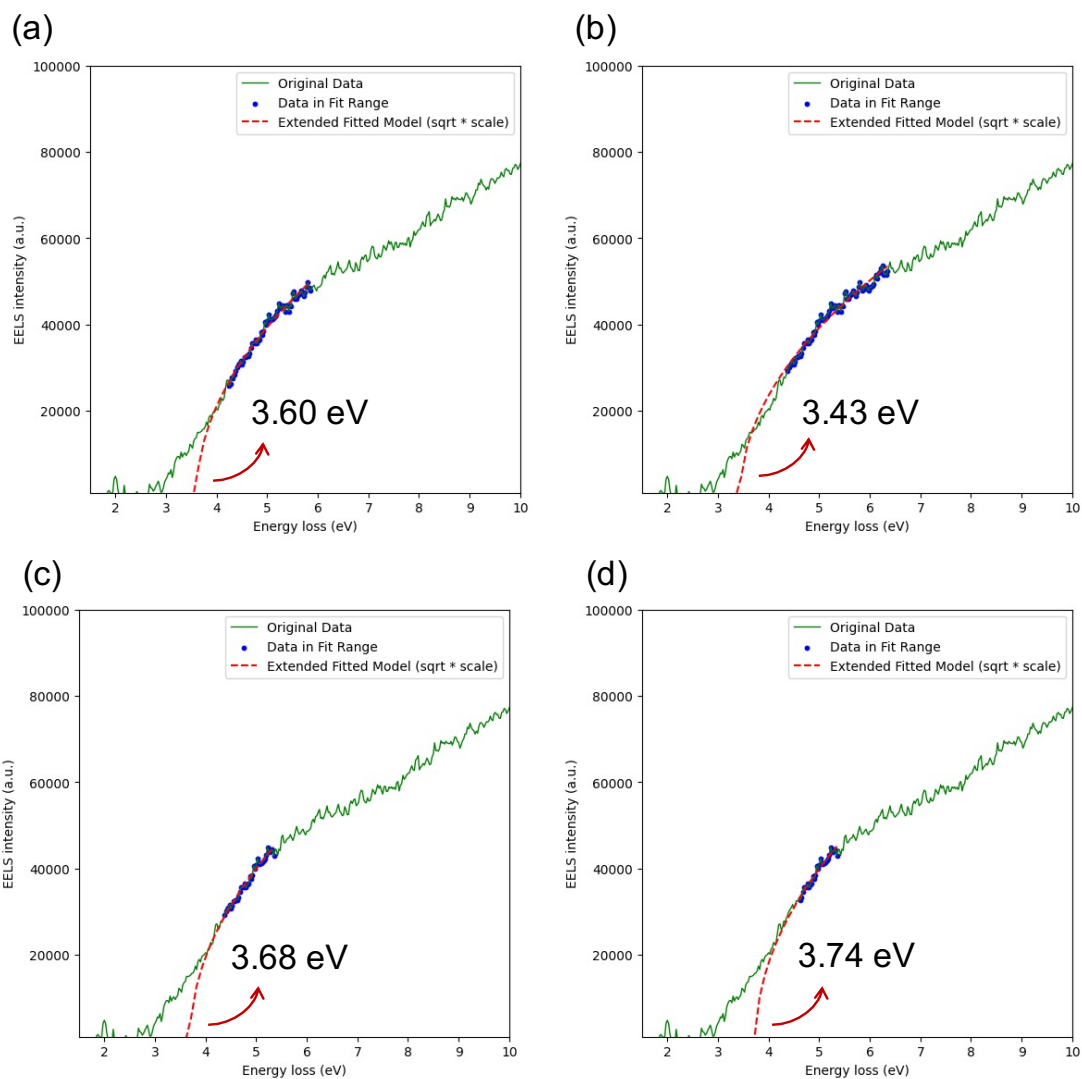

**Figure S14.** Low-loss EELS spectrum of EG/InO$_2$/SiC heterostructure with power law fitting (E-E$_g$)$^{0.5}$. Extracted absorption energy highly depends on the chosen energy window for the fit, as variations in this window yielded fit results ranging from 3.43 to 3.74 eV.

The structural modifications of InO$_2$ in 2D also impact its phonon band structure. Figure S15a demonstrates the Raman spectra taken from In, InO$_2$ and mixed In/InO$_2$ intercalated EG/SiC. Upon full oxidation, indium ULF peaks (17, 45, and 96 cm$^{-1}$) disappear (Figure S15a). On the other hand, in partially oxidized sample, in addition to the residual ULF peaks, 2D InO$_2$ peaks emerge at 295, 323, 421, and 450 cm$^{-1}$, likely due to surface enhanced Raman scattering (SERS) effect when near metallic indium. Auger spectroscopy confirms this partial oxidation; the peak for metallic indium shifts from 403 eV to 399.8 eV when fully oxidized, similar to bulk In$_2$O$_3$ (Figure S15b). Partially oxidized samples show a peak at 401.8 eV with a shoulder at 403 eV, indicating some metallic indium remains.

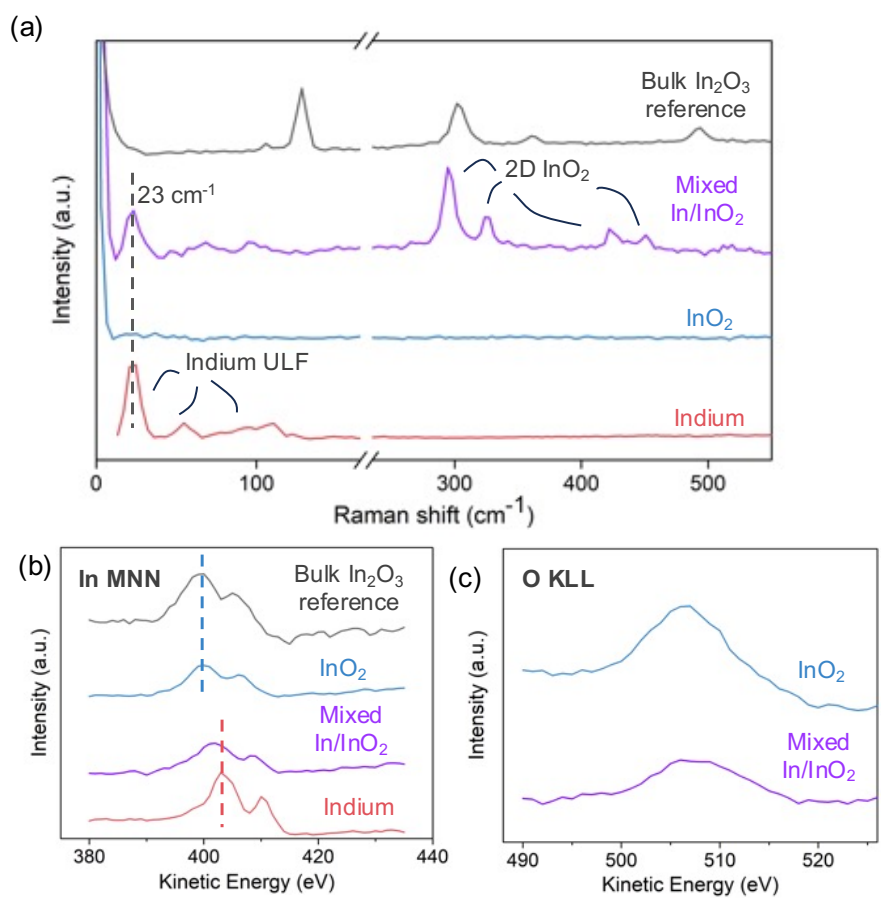

**Figure S15.** Raman (a) and Auger In MNN (b), O KLL (c) spectra taken from indium, InO$_2$, and mixed In/InO$_2$ intercalated EG/SiC and bulk In$_2$O$_3$ particles. "Mixed In/InO$_2$" sample refers to partially oxidized EG/In/SiC at 600 °C for 15 min and demonstrates 2D InO$_2$ Raman bands.

To investigate the origin of the Raman peaks observed after oxidation the phonon band structure of the EG/InO$_2$/SiC heterostructure was calculated where the structure in Figure S12c, d was used.

Before discussing our findings, we acknowledge the inherent challenges in achieving precise agreement between experimental and calculated Raman intensities, especially for resonant Raman spectra, as discussed in the literature.[46,50] Discrepancies often arise due to factors such as excitonic effects, anharmonicity, and variations in experimental conditions. Given these limitations, we interpret the calculated intensities primarily as qualitative indicators of Raman activity, confirming the presence of active modes under the specific backscattering and laser polarization conditions rather than aiming for an exact replication of experimental intensities. Figure S16 shows the Raman spectra of the 2D $InO_2$ system, comparing experimental and theoretical results. The black line represents the experimental Raman spectrum of an encapsulated 2D $InO_2$ system, with peak positions determined by fitting Lorentzian to the raw data summarized in Table S2. The green line shows the calculated Raman spectrum of an idealized structure designed to approximate the experimental sample. The green vertical bars highlight calculated Raman peaks that lie within 15 cm$^{-1}$ of the experimental peak positions, see Table S2. The red dashed line depicts the calculated Raman spectrum of bulk $In_2O_3$. While low-frequency peaks are observed experimentally in metallic In and mixed In/$InO_2$ systems, several such peaks are also present exclusively in the 2D $InO_2$ system but absent in bulk $In_2O_3$. Thus, these low-frequency modes can serve as fingerprints of the 2D nature of the system.

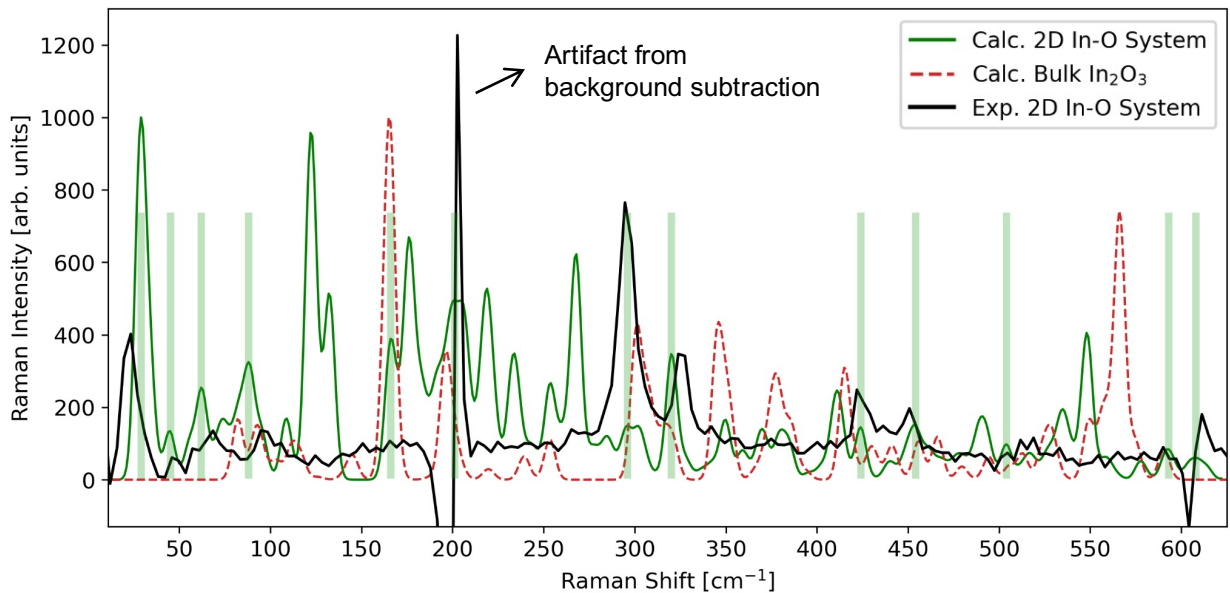

**Figure S16:** Experimental and calculated Raman spectra of the 2D $InO_2$ system, with the experimentally observed spectrum shown as a black line and the calculated spectrum as a solid green line. The red dashed line represents the calculated Raman spectrum of bulk $In_2O_3$. Green vertical bars indicate the positions of calculated Raman peaks that fall within 15 cm$^{-1}$ of the corresponding experimental peaks. The exact peak positions are summarized in Table S2.

**Table S2:** Peak positions extracted from fits of the experimental spectrum, compared to calculated peak positions within 15 cm$^{-1}$ of the experimental values. Lorentzian fits to the experimental Raman peaks were used to determine the experimental peak positions. To initialize the fitting process, the raw data was analyzed using the find_peaks function from scipy.signal, with a prominence threshold set at 3% relative to the baseline background.

| Peak label | Experiment [cm$^{-1}$] | Calculation [cm$^{-1}$] |
|:---:|:---:|:---:|
| P1 | 22.8 | 29 |
| P2 | 51 | 45 |
| P3 | 68.6 | 62 |
| P4 | 97.1 | 88 |
| P5 | 168 | 166 |
| P6 | 203 | 201 |
| P7 | 295.8 | 296 |
| P8 | 324.7 | 320 |
| P9 | 424.7 | 424 |
| P10 | 448.7 | 454 |
| P11 | 517 | 504 |
| P12 | 587.8 | 593 |
| P13 | 614.4 | 608 |

We can understand the unique 2D nature of the low-frequency peaks by visualizing the underlying atomic motions responsible for the peaks. Figure S17a illustrates the atomic dis- placements associated with the P1 Raman mode, calculated at 29 cm$^{-1}$. The structure represents an encapsulated 2D $InO_2$ system, with atoms color-coded as follows: brown (C atoms), blue (Si atoms), red (O atoms), and light purple (In atoms). The green arrows indicate the vibrational eigenvectors, showing the directions of atomic motion. This low-frequency mode primarily involves out-of-plane oscillations of the oxygen-coordinated In atoms, coupled with distortions in the underlying 2D framework. The encapsulating graphene layer exhibits minimal displacement, suggesting that the motion is localized within the $InO_2$ system. The dominance of out-of-plane motion in this mode aligns with the low-frequency Raman peaks serving as fingerprints of the 2D nature of the system as the interfacial force constants between the In-O system and the encapsulated graphene layer are being probed during this vibration. Similarly, the vibrational modes responsible for peaks P2-P4 (low-frequency modes) primarily involve atomic displacements within the $InO_2$ layer with significant out-of-plane displacement components of the In-atoms eigenvectors. Higher-frequency modes behind P5 and the following bands on the other hand, increasingly involve in-plane motions of In and other atoms including those in the graphene top or the SiC bottom layers.

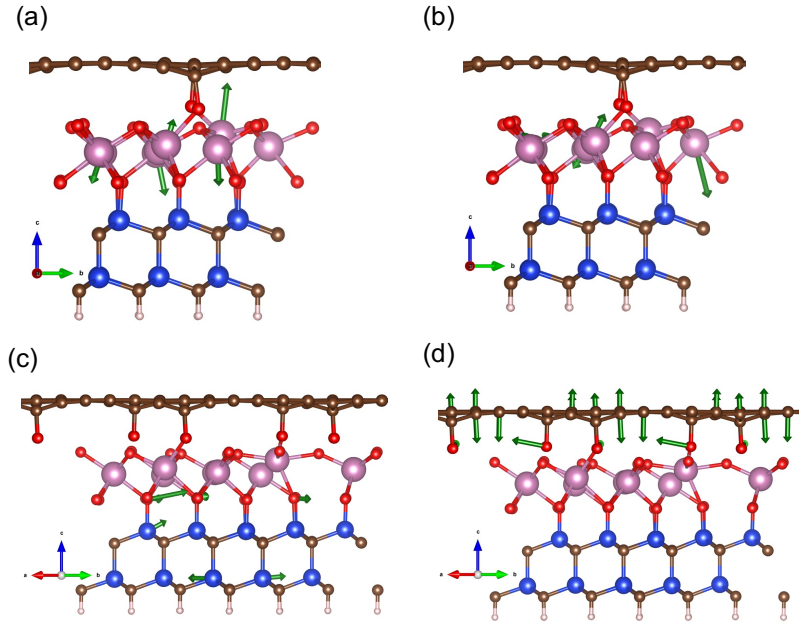

**Figure S17:** Visualization of the highest intensity phonon modes in $InO_2$, where panels (a), (b), (c), (d) demonstrate modes denoted as P1, P3, P7, P8 in Table S2.

Compared to the bulk $In_2O_3$ Raman peaks at 303, 360, and 492 cm-1,[27] 2D $InO_2$ peaks are red shifted which is attributed to the expansion of In-O bonds influenced by Si-O and C-O bonds present. Structural calculations show that the bond distance between the top oxygen and indium increases from 2.05 Å to 2.2 Å due to Gr-O bonding (Figure S18). Likewise, the presence of Si-O bonds increases the bottom O-In distance to 2.25 Å, larger than the one in bulk $In_2O_3$ (2.14 Å),[28] explaining the observed red shift in the Raman spectra due to bond expansion at the EG/SiC interface.

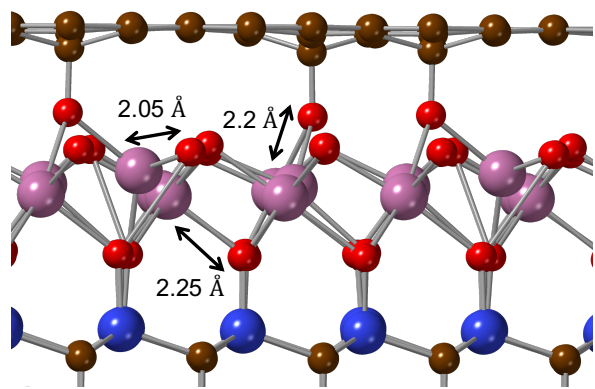

**Figure S18.** Calculated structure of InO$_2$ intercalated EG/SiC with Gr-O bonding, demonstrating In-O bond expansion compared to bulk In$_2$O$_3$ (2.14 Å) due to Gr-O and Si-O bonding. Si, C, In, and O atoms are represented by blue, brown, light purple, and red.

**Indium and InO$_2$ Intercalated EG/n-SiC Vertical Diodes**

### Device Fabrication

Indium and InO$_2$ intercalated EG/n-SiC vertical diodes were fabricated via e-beam lithography (EBL) (Figure S19). First, circular top electrode contact with 0.6 μm diameter was made with Ti/Au (5:40 nm) lift-off on graphene. Then, graphene/InO$_2$ heterostructure was etched with O$_2$ plasma in 30 seconds to minimize the device size and make sure current does not flow through non-intercalated regions which reduces the threshold voltage and rectification ratio. As an isolation layer, 30 nm Al$_2$O$_3$ was deposited via atomic layer deposition (ALD) at 150 °C using H$_2$O and trimethylaluminum (Al$_2$(CH$_3$)$_6$). Then, contacts were opened by etching Al$_2$O$_3$ in BCl$_3$ and Cl$_2$ (30 sccm:10 sccm) plasma. Ti/Au (10:120 nm) leads and pads were defined by lift-off. The bottom contact (n-SiC) was made by depositing Ti/Au (5:40 nm) on multilayer graphene grown on the C face of n-SiC. Finally, the sample was electrically connected to a gold coated AFM stainless steel disks using silver paste to facilitate probe contact. Electrical measurements were conducted with the n-SiC substrate grounded, while voltage was applied to the graphene layer.

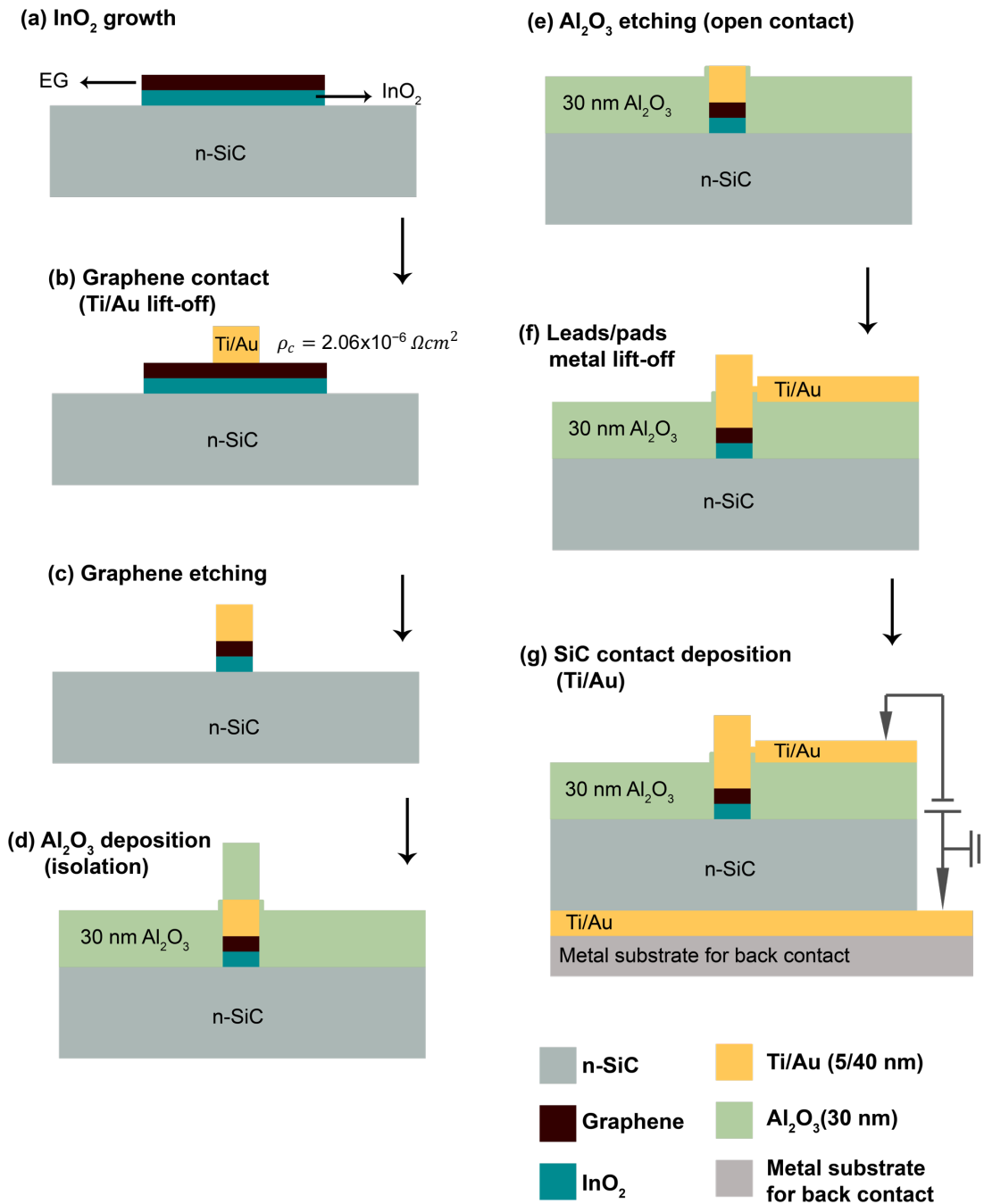

**Figure S19.** Device fabrication steps of the EG/InO$_2$/n-SiC diode using e-beam lithography. Top electrode (graphene) contact lift-off with Ti/Au (5:40 nm) (b) following InO$_2$ intercalation (a). Top contact has circular geometry with 0.6 µm diameter. Graphene and InO$_2$ etching with O$_2$ plasma (c). 30 nm Al$_2$O$_3$ deposition via ALD for isolation between metal leads and n-SiC substrate (d). Al$_2$O$_3$ etching to open the top contact using BCl$_3$/Cl$_2$ plasma (e). Ti/Au (10:120 nm) leads and pads lift-off (f). Bottom electrode (n-SiC) deposition with Ti/Au (5:40 nm) at the backside of n-SiC (g). Device has asymmetric geometry with a circular small size top contact (0.6 µm diameter) and 1x1 cm$^2$ bottom contact (whole sample).

Device fabrication recipes for both optical and e-beam lithography are given below (Table S3). Prior to spin coating, all the samples were dehydration baked at 150 °C for 3 min and then cooled down to room temperature for 2 min.

Table S3: Device fabrication recipes for EG/InO$_2$/n-SiC diode via optical and e-beam lithography.

| **Alignment Mark with SiC etching – Optical Lithography** |
|---|
| Spin LOR5A at 4K, 45 sec. Soft bake at 180 °C, 180 sec. |
| Spin SPR 955 at 3K, 45 sec. Soft bake at 105 °C, 120 min. |
| Expose via optical lithography (Heidelberg: MLA 150) with a dose of 180 mJ/cm$^2$, |
| Develop with CD-26 in 75 sec and rinse in DI-water for 60 sec. |
| Etch SiC 550 nm deep via SF$_6$ plasma. Parameters: 20mTorr, 55 sccm SF$_6$ (no O$_2$), 20 secs etch and 60 sec stabilization for 13 cycles. Etch rate: 43 nm/cycle. |
| Strip the resist in PRS 3000 at 80 °C in 30 min. Rinse with IPA for 5 min. |
| **Graphene etching (isolation etch) – Optical Lithography** |
| Spin SPR 3012 at 4K, 45 sec. Soft bake at 95 °C, 60 sec. |
| Expose via optical lithography (Heidelberg: MLA 150) with a dose of 200 mJ/cm$^2$, |
| Develop with CD-26 in 60 sec and rinse in DI-water for 60 sec. |
| Etch graphene in N$_2$ plasma for 15 sec. Antenna power: 200 W, Bias: 10W. N$_2$: 20 sccm, |
| Strip the resist in PRS 3000 at 80 °C in 30 min. Rinse with IPA for 5 min. |
| **Top Contact lift-off (Ti/Au:5/40 nm) – E-beam lithography** |
| Spin MMA EL6 (150 nm) at 4K, 45 sec. Soft bake at 150 °C, 90 sec. |
| Spin PMMA 950 A3 (180 nm) at 4K, 45 sec. Soft bake at 180 °C, 90 sec. |
| Expose via EBL (Raith: EBPG 5200 Vistec) - 330 μC/cm$^2$ dose. |
| Develop with MIBK/IPA (1:1) for 60 sec and rinse in IPA for 45 sec. |
| Deposit 5/40 nm thick Ti/Au via e-beam evaporation. Dep rate: 0.5 Å/sec for Ti and 2 Å/sec for Au. |
| Lift-off with acetone (50 °C for 15 min) and then PRS 3000 (80 °C for 30 min). Rinse with IPA for 5 min. |
| **Al$_2$O$_3$ etching – E-beam lithography** |
| Spin Zep (undiluted) at 2.5K, 45 sec. Soft bake at 180 °C, 180 sec. |
| Expose via EBL (Raith: EBPG 5200 Vistec) - 375 μC/cm$^2$ dose. |
| Develop with N-amyl acetate for 180 sec and rinse in IPA for 60 sec. |
| Etch Al$_2$O$_3$ via BCl$_3$/Cl$_2$ plasma (5 °C, 30 sccm BCl$_3$, 10 sccm Cl$_2$, 60 secs (4 repetition). Etch rate at 5 °C is 6 Å/sec. |
| Strip the resist in PRS 3000 at 80 °C in 30 min. Rinse with IPA for 5 min. |
| **Metal pads/leads lift-off (Ti/Au:10/120 nm) – E-beam lithography** |
| Spin MMA EL11 at 4K, 45 sec. Soft bake at 150 °C, 90 sec. |
| Spin PMMA 950 A3 at 4K, 45 sec. Soft bake at 180 °C, 90 sec. |
| Expose via EBL (Raith: EBPG 5200 Vistec) - 375 μC/cm$^2$ dose. |
| Develop with MIBK/IPA (1:1) for 60 sec and rinse in IPA for 45 sec. |
| Deposit 10/120 nm thick Ti/Au via e-beam evaporation. Dep rate: 0.5 Å/sec for Ti and 2 Å/sec for Au. |
| Lift-off with acetone (50 °C for 15 min) and then PRS 3000 (80 °C for 30 min). Rinse with IPA for 5 min. |
| **Bottom contact (Ti/Au:5/40 nm) – No lithography** |
| Ti/Au (5/40 nm) deposition at the multilayer graphene grown on C face of SiC. |

### Graphene Contact Resistance

We first examined the electrical properties of the graphene/Ti/Au interface to assess the contact resistance of the top electrode. Figure S20a displays room temperature resistance measurements taken from graphene, contacted with Ti/Au (5:40 nm) on insulating 6H-SiC substrate using circular transfer length method (CTLM) with changing electrode spacing d (channel length in Figure S20b). The measurements were taken using four-point probe technique to eliminate resistance originating from the probes. The results confirmed ohmic I-V characteristics across all voltage levels. A geometry correction for CTLM has properly been carried out, following the procedure in Ref[51]. Each point in Figure S20b represents the slope of a linear I-V curve. The linear relationship observed with distance allowed us to calculate the contact resistivity ($\rho_C = 2.06 \times 10^{-6}$ $\Omega cm^2$), sheet resistance ($R_{sh} = 0.14$ $\Omega$/sq) and the transfer length ($L_c = 0.39$ μm), based on the Equations S1-S3, where $R_T$ is total resistance measured via four-point probe, $R_c$ is graphene/Ti/Au contact resistance, $R_{sh}$ is sheet resistance, $l$ is contact spacing, and $W$ is contact area, $L_T$ is transfer length.

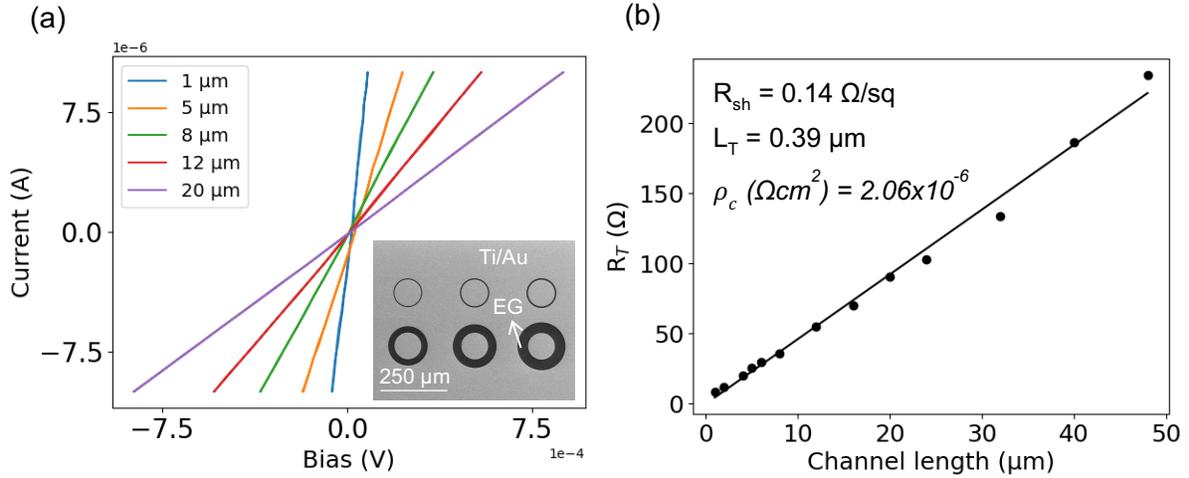

**Figure S20:** Electrical properties of graphene/Ti/Au contact using CTLM. Current-voltage curves taken from Ti/Au contacted graphene with various contact spacing, verifying linear relationship (a). Geometry corrected $R_T$ as a function of contact spacing $l$, demonstrating linear relationship (b). Contact resistivity of $2.06 \times 10^{-6}$ $\Omega cm^2$ is extracted from the intercept of (b).

$$R_T = 2R_c + R_{sh} \times \frac{l}{W} \qquad \text{Equation S1}$$

$$R_T = \frac{R_{sh}}{W}(l + 2L_T) \qquad \text{Equation S2}$$

$$\rho_c = R_{sh} \times L_T^2 \qquad \text{Equation S3}$$

For the bottom electrode we contacted n-SiC from the C face (000-1) by depositing Ti/Au (5/40 nm) via e-beam evaporation without any lithography step (Figure S21a). A common approach to form low resistance ohmic contact to n-SiC is to deposit Ni and anneal at temperatures >1000 °C.[52] As annealing the sample at such high temperatures would lead to deintercalation of 2D oxide, we used the multilayer graphene grown on the C face of SiC to form ohmic contact to n-SiC with Ti/Au. When EG is grown via Si sublimation from SiC, graphene grows on all surfaces, including the backside of the SiC wafer (000-1). This interfacial graphene layer helps for the formation of ohmic contact between n-SiC and metal contact.[10] Figure S21b demonstrates I-V curve taken from Ti/Au contacted as-grown EG/n-SiC (contacted both sides), verifying ohmic conduction for vertical transport (125 Ω). Hence, observed barrier in diode measurements with intercalated structures is attributed to the modulations at the interface.

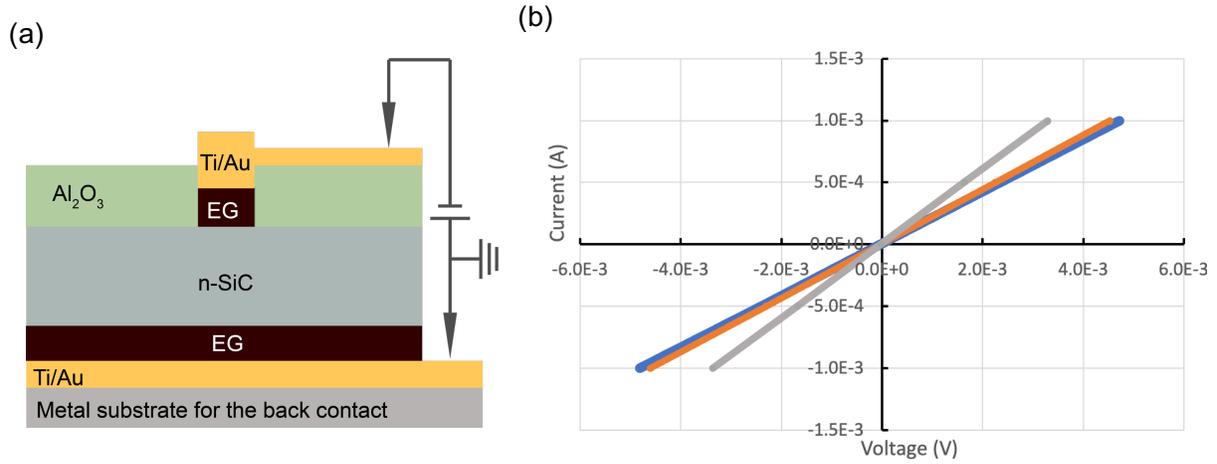

**Figure S21**. Two-terminal vertical device measurements with as-grown graphene/n-SiC contacted with Ti/Au. (a) showing the schematic for the measurement system. (b) I-V curves taken from Ti/Au contacted as-grown EG/n-SiC two terminal devices showing ohmic conduction.

**EG/InO$_2$/n-SiC Vertical Diode Measurements**

Temperature dependent I-V (10 K – 400 K) was conducted on EG/InO$_2$/n-SiC MOS-based Schottky diode to understand the vertical conduction mechanism and extract the barrier height, $\varphi_B$. The fit of the reverse bias current at 10 K with Fowler Nordheim Tunnelling (FNT) demonstrates a linear relationship (ln (I/V$^2$) vs 1/V), verifying tunneling as dominant conduction mechanism for <150 K (Figure S22a). In the forward bias, the I-V curve is examined with the thermionic emission (TE) model, based on the Equations 1-4 in the main text (see methods). After extracting the reverse saturation current $I_o$ from $ln(I)$ vs $V$ plot (inset in Figure 4b), Richardson plot has been drawn for 320 K – 400 K (Figure S22b), where Richardson Constant ($A^*$) and barrier height ($\varphi_B$) is extracted as 139 Acm$^{-2}$K$^{-2}$ and 0.19 eV, respectively. Although extracted $A^*$ is close to the theoretical Richardson constant of 4H-SiC (146 Acm$^{-2}$K$^{-2}$), the barrier height is lower than expected.

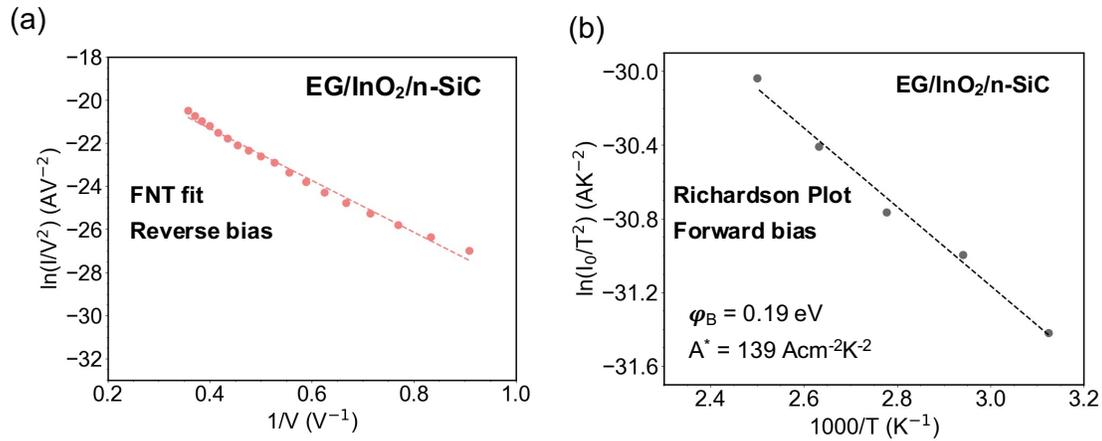

**Figure S22:** Fowler-Nordheim Tunneling (FNT) fit of the reverse bias current for EG/InO$_2$/n-SiC vertical diode at 10 K, demonstrating linear relationship. Richardson plot for the forward bias current at 320 K – 400 K, where A$^*$ and $\varphi_B$ are extracted as 139 Acm$^{-2}$K$^{-2}$ and 0.19 eV, respectively.

Alternatively, the barrier height at each temperature can be extracted from Equation 2 (main text) by using theoretical $A^*$ for 4H-SiC and device area (0.28 μm$^2$) for 320 K – 400 K. As shown in Table S4, with higher temperatures, the barrier height increases from 0.39 eV to 0.49 eV and ideality factor reduces from 3.87 to 1.76. Such a temperature dependent behavior of $\varphi_B$ and $n$ is attributed to the inhomogeneity of the barrier height. At low temperature, electrons without sufficient energy can only surmount patches with lower Schottky barrier. As the temperature increases, more and more electrons gain sufficient energy to overcome higher barrier, leading to the measurement of higher apparent Schottky barrier height.[31]

**Table S4.** Reverse saturation current density, $J_0$, barrier height, $\varphi_B$, and ideality factor, $n$, extracted from Equation 4 in the main text for the forward bias current for EG/InO$_2$/n-SiC device. Increase in $\varphi_B$ and reduction in $n$ with temperature indicate barrier inhomogeneities at the EG/InO$_2$/n-SiC interface.

| Temperature (K) | $J_0$ | Barrier height (eV) | Ideality factor (n) |
|---|---|---|---|
| 320 | 10.49 | 0.39 | 3.87 |
| 340 | 5.89 | 0.44 | 2.82 |
| 360 | 8.94 | 0.45 | 2.18 |
| 380 | 20.29 | 0.45 | 2.16 |
| 400 | 14.88 | 0.49 | 1.76 |

The presence of barrier inhomogeneities is directly observed by conductive-AFM (C-AFM) measurements in both current mapping and I-V curves (Figure S23). Figure S23b presents several I-V curves taken from different regions on the EG/InO$_2$/n-SiC sample, where reverse $V_{th}$ changes between -4 V and -6 V. Additionally, in the C-AFM current mapping (Figure S23a), high conductivity regions are present, indicating barrier height inhomogeneities.

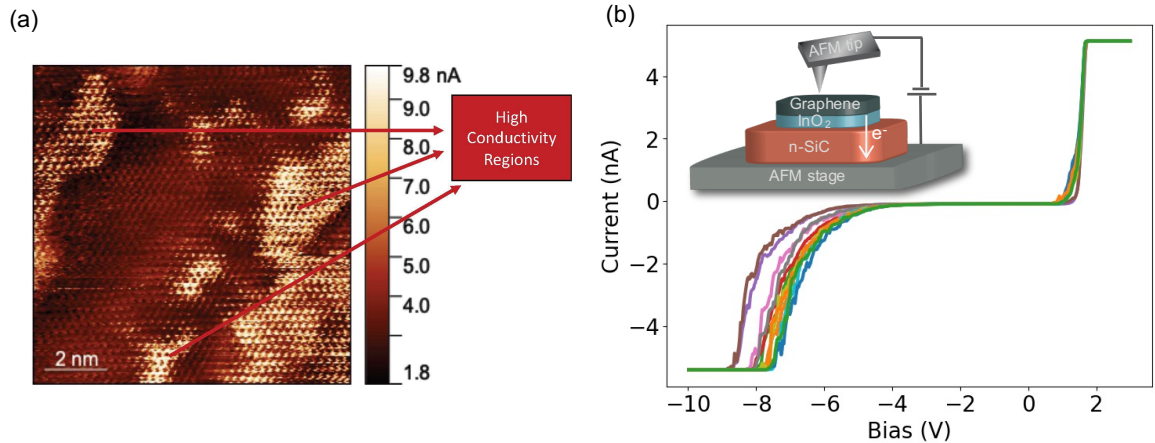

**Figure S23.** Nanoscale electrical characterization of the EG/InO$_2$/n-SiC vertical diode via C-AFM. High conductivity regions are observed in the atomic resolution current map in (a), indicating inhomogeneities in the barrier height. (b) Several I-V curves acquired from different regions on the sample, verifying variation in reverse leakage current with $V_{th}$ changing between -4 V and -6 V.